\documentclass[12pt]{article}
\usepackage{epsf}

\newcommand{\bmat}{\left(\begin{array}}
\newcommand{\emat}{\end{array}\right)}

\def\yzero{\smash{\hbox{$y\kern-4pt\raise1pt\hbox{${}^\circ$}$}}}

\def\g{\gamma}

\def\beq{\begin{equation}}
\def\eeq{\end{equation}}
\def\beqa{\begin{eqnarray}}
\def\eeqa{\end{eqnarray}}

\def\th{\theta}

\def\-{\hphantom{-}}
\def\ov{\overline}
\def\s2{\frac{1}{\sqrt2}}

\def\beq{\begin{equation}}
\def\eeq{\end{equation}}
\def\beqa{\begin{eqnarray}}
\def\eeqa{\end{eqnarray}}

\def\Tr{{\rm Tr \,}}
\def\diag{{\rm diag \,}}
\def\IF{\relax{\rm I\kern-.18em F}}
\def\II{\relax{\rm I\kern-.18em I}}
\def\IP{\relax{\rm I\kern-.18em P}}
\def\IC{\relax\hbox{\kern.25em$\inbar\kern-.3em{\rm C}$}}
\def\IR{\relax{\rm I\kern-.18em R}}

\def\Dsl{\,\raise.15ex\hbox{/}\mkern-13.5mu D} 
\def\IZ{Z\kern-.4em  Z}
\def\id{{\rm I}}

\def\IC{\relax{\rm I\kern-.48em C}}
\def\NN{{\cal N}}
\def\BL{B{\rm -}L}

\newcommand{\drawsquare}[2]{\hbox{%
\rule{#2pt}{#1pt}\hskip-#2pt
\rule{#1pt}{#2pt}\hskip-#1pt
\rule[#1pt]{#1pt}{#2pt}}\rule[#1pt]{#2pt}{#2pt}\hskip-#2pt
\rule{#2pt}{#1pt}}

\newcommand{\fund}{\raisebox{-.5pt}{\drawsquare{6.5}{0.4}}}
\newcommand{\antifund}{\overline{\fund}}

\catcode`\@=11   
\newdimen\@rotdimen
\newbox\@rotbox  

\def\@vspec#1{\special{ps:#1}}
\def\@rotstart#1{\@vspec{gsave currentpoint currentpoint translate
   #1 neg exch neg exch translate}}
\def\@rotfinish{\@vspec{currentpoint grestore moveto}}
\def\@rotr#1{\@rotdimen=\ht#1\advance\@rotdimen by\dp#1%
   \hbox to\@rotdimen{\hskip\ht#1\vbox to\wd#1{\@rotstart{90 rotate}%
   \box#1\vss}\hss}\@rotfinish}
\def\@rotl#1{\@rotdimen=\ht#1\advance\@rotdimen by\dp#1%
   \hbox to\@rotdimen{\vbox to\wd#1{\vskip\wd#1\@rotstart{270 rotate}%
   \box#1\vss}\hss}\@rotfinish}%
\def\@rotu#1{\@rotdimen=\ht#1\advance\@rotdimen by\dp#1%
   \hbox to\wd#1{\hskip\wd#1\vbox to\@rotdimen{\vskip\@rotdimen
   \@rotstart{-1 dup scale}\box#1\vss}\hss}\@rotfinish}%
\def\@rotf#1{\hbox to\wd#1{\hskip\wd#1\@rotstart{-1 1 scale}%
   \box#1\hss}\@rotfinish}%
\def\rotate{\@ifnextchar[{\@rotate}{\@rotate[l]}}
\def\@rotate[#1]#2{\setbox\@rotbox=\hbox{#2}\@nameuse{@rot#1}\@rotbox}

\catcode`\@=12

\topmargin
-1.5cm
\textwidth
15.5cm
\textheight
23.5cm
\oddsidemargin
0.7cm
\evensidemargin
1.2cm

\begin{document}

\makeatletter
\@addtoreset{equation}{section}
\makeatother
\renewcommand{\theequation}{\thesection.\arabic{equation}}
\pagestyle{empty}
\rightline{ IFT-UAM/CSIC-07-57}
\rightline{ CERN-PH-TH/2007-215}
\rightline{\tt hep-th/yymmnnn}
\vspace{1.5cm}
\begin{center}
\LARGE{ Instanton Induced Open String\\ 
Superpotentials and Branes at Singularities   
\\[10mm]}
\large{ L.E. Ib\'a\~nez$^{1}$ and A. M. Uranga$^{2}$ \\[6mm]}
\small{
${}^{1}$ Departamento de F\'{\i}sica Te\'orica C-XI
and Instituto de F\'{\i}sica Te\'orica UAM-CSIC,\\[-0.3em]
Universidad Aut\'onoma de Madrid,
Cantoblanco, 28049 Madrid, Spain \\[2mm] and 
\\[2mm]
${}^{2}$ PH-TH Division, CERN \\
CH-1211 Geneva 23, Switzerland\\
(On leave from IFT-UAM/CSIC, Madrid)\\
[7mm]}
\small{\bf Abstract} \\[12mm]
\end{center}
\begin{center}
\begin{minipage}[h]{16.0cm}
We study different aspects of the non-perturbative superpotentials induced by Euclidean $E3$-branes on
systems of $D3$/$D7$-branes located  at Abelian orbifold singularities. We discuss in detail how the induced couplings are consistent with the $U(1)$ symmetries carried  by the $D3/D7$ branes.
We construct different compact and non-compact examples, and show 
phenomenologically relevant couplings like $\mu$-terms or certain  Yukawa 
couplings generated by these E3 instantons.  Some other novel effects are described. We show an example where $E3$ instantons combine with standard gauge instantons to yield new multi-instanton effects contributing to superpotential,  along the lines of ref.\cite{geu}. In the case of non-SUSY $Z_N$ tachyon-free singularities it is shown how $E3$-instantons give rise to non perturbative scalar couplings including exponentially suppressed scalar bilinears.

\end{minipage}
\end{center}
\newpage
\setcounter{page}{1}
\pagestyle{plain}
\renewcommand{\thefootnote}{\arabic{footnote}}
\setcounter{footnote}{0}


\section{Introduction}

Euclidean brane instantons (see e.g. \cite{instantons})
 provide the leading non-perturbative contribution to 
several important quantities in string compactifications. 
For instance D-brane instantons in type II compactifications
 (or F/M-theory duals) have led to (in some cases very explicit)
 proposals to lift 4d moduli which are otherwise flat 
directions of the theory in the classical supergravity 
approximation. In addition, instantons arising from 
euclidean D$p$-branes wrapped on $(p+1)$-cycles (henceforth E$p$-branes),
 intersecting the 4d spacefilling D-branes yielding the gauge group,
 provide the leading contribution to perturbatively 
forbidden superpotential couplings of the relevant 4d 
effective gauge theory \cite{bcw,iu} 
(see also \cite{ganor,Haack:2006cy,fkgs}
 and \cite{ablps,isu,Cvetic:2007ku,
abls,abls2,aim,bclrw,cw} for related applications).

One of the main problems  in improving 
the understanding and  potential applications of these
 instantons lies in the difficulty in constructing 
explicit models where the instantons have the appropriate 
number of zero modes to lead to contributions to the 
non-perturbative superpotential. In particular,
 the non-vanishing of such contributions depends on 
the presence of non-trivial couplings for all fermion 
zero modes (charged or uncharged under the 4d gauge group) 
except the two 4d $N=1$ Goldstinos. 
These couplings are difficult to evaluate in general 
Calabi-Yau compactifications, or even for exactly 
solvable but non free-field CFTs. Such couplings 
can however be explicitly discussed in compactifications 
on toroidal orientifolds (see e.g.  \cite{bk,bfm} for explicit examples),
 or for instantons from branes wrapped on compact 
cycles in systems of D-branes at singularities
 (see e.g. \cite{fkgs,abfk,abflp,Franco:2007ii,ak,aks}).

In this paper we focus on instanton effects from 
euclidean branes wrapped on non-compact cycles 
in local models of D3/D7-branes at singularities. 
More specifically we consider  effects from E3-brane 
instantons wrapped on holomorphic 4-cycles passing though 
the singular point (and thus intersecting the 4d spacefilling D-brane 
system). Such instanton effects become physical when the local model 
is embedded in a full-fledged compactification, 
but many of its properties (and in particular the structure
 of 4d charged fields involved in the effective 
vertex they produce) depend crucially only on the 
local model (plus some mild assumption about 
behaviour at infinity). Note that E3-brane instantons 
may not be the only instanton effects in such global 
compactifications, but they are the most generic, 
in the following sense. For instance, E5-branes wrapped on 
the whole internal space, and carrying stable 
holomorphic world-volume gauge bundles, 
provide additional instantons. However, note that in 
general such instantons are BPS only at particular loci 
in moduli space, away from which they cross lines of marginal 
stability 
We thus stick to the simpler situation of E3-brane 
instantons, and to compact examples where they are 
BPS all over the moduli space.

We moreover concentrate on the case of abelian 
orbifold singularities, given the simple and powerful 
CFT description of the instanton zero modes and 
their interactions, and the relative ease to embed 
such models (at least for low order orbifolds) 
in toroidal orientifold compactifications. Similar 
ideas can be applied for other non-orbifold but 
toric singularities, using dimer diagram techniques, 
as we sketch in an appendix.

A further motivation to consider this kind of system 
is phenomenological. Indeed, one of the most attractive 
possibilities to embed the Standard Model in string theory 
is via systems of D-branes at singularities, since they 
naturally lead to world-volume chiral gauge theories.
 In fact, several realizations of semirealistic models
 have been proposed
\cite{aiqu,bjl,gerardo,bkl,kp,vw}. 
It is therefore a natural question to consider the
 structure of field theory operators that can be induced 
by instanton effects in this setup. As we explain in next
 Section, the fact that the most promising singularities 
leading to semi-realistic models are not orientifold 
singularities implies that the only D-brane instantons 
contributing to the superpotential are (except for brane 
instantons with gauge theory interpretation) those wrapped 
on non-compact cycles passing through the singularity, 
and through orientifold planes away from the latter.

We provide the general formalism to study such effects 
for general supersymmetric abelian orbifold singularities, 
and illustrate it with a set of explicit examples 
leading to a rich pattern of physical phenomena 
(supersymmetry breaking, generation of mass terms,...) 
and interesting superpotential couplings forbidden in 
perturbation theory (quark and/or lepton Yukawa couplings,
 Higgs $\mu$-terms, etc). In addition,
 we also describe possible effects of brane 
instantons in (non-tachyonic) non-supersymmetric orbifold models,
 which have not been described in the literature, 
and argue that they generate potentials for the 4d charged scalars. 

There is another interesting aspect to this class of
$E3$ instantons. It has been recently pointed out in \cite{geu} 
that instantons with additional neutral fermion zero modes
may still contribute to a non-perturbative superpotential
if the extra zero modes are lifted by another instanton. In particular we find 
that $E3$ instantons 
may combine with standard $E(-1)$ gauge instantons to
 induce novel superpotential couplings and
provide an explicit compact $\IZ_3$ orientifold example 
in which this phenomenon takes place.

The paper is organized as follows. 
In Section \ref{ethree} we describe the general 
features of E3-brane instantons for systems 
of D3/D7-branes at $\IC^3/\IZ_N$ local singularities, 
in particular we provide the explicit description
 of the structure of their charged and neutral zero modes.
 In Section \ref{action} we discuss the instanton action
 and its coupling to RR fields, and show the gauge 
invariance of the instanton effective vertex under gauge 
transformations of the 4d spacefilling gauge D-branes.
 In Section \ref{examples} we provide explicit
 examples of non-compact and compact models 
and interesting instanton effects, including 
 examples of phenomenologically interesting 
couplings in semi-realistic examples of \cite{aiqu}. 
We also describe there a realization of the new phenomenon
of superpotential contributions from multi-instantons \cite{geu}. We also 
discuss  the  generation of Fayet SUSY breaking 
induced by $E3$ instantons.
Section \ref{nonsusy} describes the novel situation 
of instanton effects  for systems of D-branes at 
(non-tachyonic) non-supersymmetric singularities. 
Section \ref{final} contains our final remarks.
 Appendix \ref{singus} describes the computation 
of the 4d field theory on D3/D7-brane systems at
 orbifold singularities. In appendix \ref{zseven} 
we present a compact $\IZ_7$ toroidal orientifold 
example. Appendix \ref{dimers} sketches the generalization 
of the instanton effects from E3-branes on 4-cycles 
to general toric singularities, based on dimer diagram techniques.

\section{Euclidean E3-brane instantons at $\IC^3/\IZ_N$ singularities}
\label{ethree}

\subsection{Generalities}
\label{generalities}

In this article we consider euclidean D-brane instanton effects on systems of $D3$- and $D7$-branes at $\IR^6/\IZ_N$ singularities. We summarize the basic formalism to compute the spectrum and interactions on the world-volume of $D3$- and $D7$- branes at Abelian singularities in the appendix (see \cite{aiqu} for more details and references).
We will mostly concentrate on   the case of  supersymmetric singularities, and on euclidean instantons contributing to the non-perturbative superpotential  \footnote{We denote E$p$ a $(p+1)$-dimensional euclidean D-brane wrapped on a $(p+1)$-cycle.}
. Later on we will briefly discuss the case of euclidean instantons at  $\NN=0$ non-supersymmetric singularities, and comment on the resulting non-perturbative interactions.

In order to contribute to the superpotential, one has to focus on BPS D-branes. Hence
there are two classes of euclidean D-brane instantons which can contribute to the superpotential in these systems, $E(-1)$ and $E3$ Euclidean branes \footnote{Here we refer to the dimension of the brane as described in the parent space. Since we work with fractional branes, in the quotient space they in general correspond, in the geometric large volume limit, to E3/E1/E$(-1)$ bound states.}.
When such euclidean branes wrap the same cycle and have the same Chan-Paton transformation properties as some of the 4d-spacefilling branes in the background, they can be interpreted as gauge instantons. Otherwise they correspond to genuine stringy effects, without field theory interpretation, which we hence dub stringy instantons. We will be most interested in the latter.

$E(-1)$ instanton induced superpotentials  have  been discussed mostly for the case of the 
conifold or the orbifolded conifold in \cite{fkgs,abfk,abflp,ak,aks} (these can be regarded as 
bound states of E3-E1-E(-1) on compact cycles in the blownup limit).
As emphasized in \cite{abflp,bfm,isu}, both $E(-1)$ and $E3$ 
stringy instantons on Calabi-Yau compactifications generically have at least four neutral 
fermionic zero modes, and thus cannot contribute to the superpotential. The two extra fermion 
zero modes are Goldstinos of an accidental enhancement of $N=1$ to $N=2$ in the E$p$-E$p$ open 
string sector. A simple way (and seemingly the only one in perturbative models) to reduce the 
number of universal zero modes to two is to consider instantons mapped to themselves under the 
orientifold action. For a $O(1)$ CP symmetry one obtains the required number of zero modes in 
that sector (clearly, there may be additional fermion zero modes in other sectors).

Focusing on systems of D-branes at singularities, for a $E(-1)$ instanton to induce a non-perturbative superpotential term involving charged chiral multiplets from the D-branes at the singularity, the $E(-1)$ must also sit at the singularity,  which thus must also be fixed under the orientifold action (thus, it is an orientifolded singularity).
This is the case considered e.g. in \cite{fkgs,abfk,aks}. In orbifold language, 
the instanton corresponds to a fractional E$(-1)$-brane, with Chan-Paton phases not associated to any of the background D3-branes (i.e. corresponding to an unoccupied node in the quiver). Geometrically, it corresponds to Euclidean $E1$-brane wrapping a collapsed 2-cycle at the singularity on which no 4d-spacefilling brane wraps.
These instantons give rise generically to genuine stringy effects .

Non-perturbative effects have been suggested to play a key role in 
semi-realistic string models of particle physics, 
in order to generate interesting couplings which are absent in perturbation theory,
 due to perturbatively exact global $U(1)$ symmetries
\cite{bcw,iu,isu}.
There exist systems of D3/D7-branes at singularities leading to such semi-realistic 
models \cite{aiqu,bjl,gerardo,kp,bkl,vw}. It would be interesting to study the 
possible appearance of non-perturbative effects from stringy instantons
in this class of models.
 However, such models are obtained for branes systems sitting at orbifold (not orientifold) singularities.
Indeed, as argued in \cite{aiqu}, models from $D3$-branes sitting at orientifold 
singularities suffer from a generic difficulty in yielding realistic spectra. The problem arises because the orientifold projection removes from the spectrum the diagonal $U(1)$ which is always anomaly free and is crucial to obtain correct hypercharge assignments.
This means that in the class of realistic models based on branes at singularities not fixed under the orientifold action, $E(-1)$ instantons can never give rise to non-perturbative superpotentials involving phenomenologically relevant chiral fields like those of
quarks and leptons.  

This caveat is nicely avoided by E3-brane instantons, wrapping a 4-cycle passing through the singularity. The E3-brane can be mapped to itself by an orientifold action leading to orientifold planes away from the relevant singularity. The orientifold will thus map the relevant singularity to a mirror image singularity. The net effect is that the system at the singularity is insensitive to the orientifold action and can reproduce one of the semi-realistic models of \cite{aiqu,bjl,gerardo,vw}. The instanton is fixed under the orientifold action, and can have $O(1)$ CP symmetry and lead to just two fermion zero modes. Since it also intersects the D-branes at the singularity, the corresponding superpotential can lead to interesting SM operators.

From the phenomenological perspective this is one of the motivations to study 
the effects of $E3$  rather than $E(-1)$  instantons. 
Nevertheless, we will keep an open mind and consider examples of non-perturbative 
effects on systems of D-branes at orientifold singularities as well, even though 
they are less  promising from the model building point of view.

It is interesting to consider the above kind of configurations
 from the viewpoint of the local physics near 
the (non-orientifold) singularity. We have a D3/D7-brane system at a non-orientifold singularity, and an euclidean E3-instanton wrapped on a non-compact 4-cycle.
On the E3-E3 open string spectrum, there are four fermion fields from the 
accidental $N=2$ susy in this sector. However, these fields propagate on the non-compact 4-cycle of the instanton. The existence or not of fermion zero modes for the spacetime instanton is determined, from the local model viewpoint, by the boundary conditions at infinity for these fields. For E3-branes with $O(1)$ CP symmetry, the boundary condition for two of the fermion fields is that they vanish at infinity, and thus removes their corresponding zero mode.

Given their simplicity and their general applicability, it is thus convenient to study first the local models. As is clear form the above, one should nevertheless be careful with the discussion of modes supported on non-compact cycles, since the corresponding zero mode spectrum is sensitive to boundary conditions at infinity.

We thus concentrate on the effects coming from $E3$ instantons (see also \cite{bk,bfm}).
We are going to consider euclidean 3-branes $E3^r$  located at a  $\IR^6/\IZ_N$ 
singularity and wrapping a 4-cycle $\Sigma_4^r$ transverse to the D3-branes. Our 
analysis here will be local and we will assume that eventually the singularity is embedded into a compact manifold so that the action of the instantons is finite and the $D7^s$ branes give rise to physical gauge bosons and not merely to a flavour symmetry. We also assume that the E3 branes do not touch further branes which could give rise to extra zero modes to be included in the analysis. Compact examples  with these characteristics will be provided  later on.

We consider  instantons $E3^r$, r=1,2,3 which wrap a 4-cycle transverse to the $r^{th}$ complex plane (thus defined by $z_r=0$). The  CP factors for these $E3^r$ will be of the form (see the appendix for notation)
\beqa
\label{gamma7}
\gamma_{\theta,E3^r} = \diag (\id_{v_0^r}, e^{2\pi i/N} \id_{v_1^r},\ldots,
e^{2\pi i(N-1)/N} \id_{v_{N-1}^r})
\eeqa
up to an overall phase which depends on the existence or not of vector structure \cite{blpssw}.
In order to get the correct number of neutral zero modes for the instantons,
eventually we will be interested in an orientifold projection giving
rise to an $O(1)$ CP symmetry. Since orientifolds act by conjugating the Chan-Paton phases (see 
\cite{Uranga:1999mb,Feng:2001rh} for a few $\IC^2/\IZ_N$ exceptional cases where it does not) this is obtained for 
$v_0^r=1$, $v_i^r=0$ with  $i\not= 0$. This already implies that appropriate instantons exist 
only in models with vector structure, where the overall phase mentioned above is unity and the 
Chan-Paton matrix is exactly as in (\ref{gamma7}). This already excludes a large class of 
models, in particular most of the even order orbifolds (compact even order orbifolds are 
difficult to construct due to the very stringent RR tadpole cancellation conditions
\cite{afiv}, see however \cite{kr,ru}).

It is easy to show that for orbifolds acting on all three complex planes (i.e. not of the form 
$\IC^2/\IZ_N\times \IC$, the E3-E3 open string sector reduces to the universal sector of four bosonic zero modes (the 4d translational Goldstones associated to the instanton position in 4d Minkowski space) and 2 fermion zero modes for $O(1)$ instantons. As discussed in \cite{isu}, for $Sp(2)$ instantons there are 6 fermion zero modes (transforming as two triplets of the instanton gauge group symmetry) and for $U(1)$ instantons there are 4. Since the possible mechanisms to lift the additional zero modes in the latter cases require ingredients beyond those in our configurations, we will focus on the case of $O(1)$ instantons.

\begin{figure}
\epsfysize=6cm
\begin{center}
\leavevmode
\epsffile{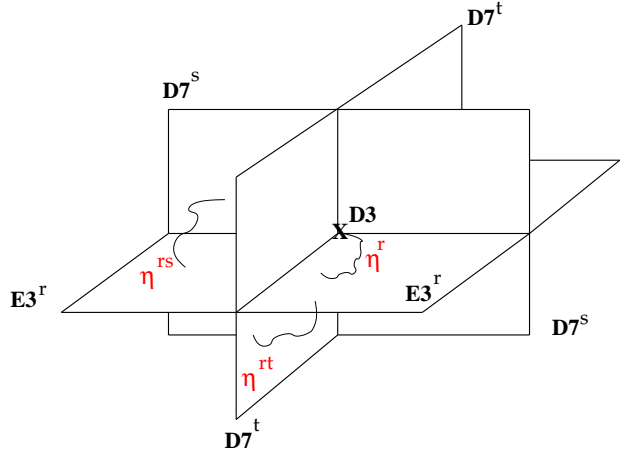}
\end{center}
\caption{Pictorial view of the instanton $E3^r$ and $D7^{s,t}$-branes
going through a singularity where a stack  of $D3$-branes are located.
The  $E3^r-D3$ and $E3^r-D7^{s,t}$ fermionic zero modes ($\eta^r$
and $\eta^{rs}$,$\eta^{rt}$ respectively) are also shown.}
\label{instanton4}
\end{figure}

Let us now discuss in turn the different types of charged fermionic zero modes which will appear in a configuration with both D3- and D7$^s$-branes. For the CP matrices of these branes we  use the notation defined in the appendix. A pictorial view of the zero modes is given in fig.(\ref{instanton4}).

\subsection{E3-D3 instanton zero modes}

This sector is fully localized in the non-compact orbifold dimensions, so it is insensitive to boundary conditions at infinity.

Open strings in the $E3^r$-$D3$ sector give rise to no bosonic zero modes.
This is due to the fact that there are altogether 8 space-time components
with mixed $DN$ boundary conditions, which lift up the zero energy of the
bosonic states. On the other hand there is a fermionic state from the 
R-state in the complex direction with DD boundary conditions. The 
gauge quantum numbers and multiplicities of these fermions is totally
analogous to the ones of the $D7^r$-$D3$ states (see eq.(\ref{specsusy})
 in the appendix), i.e. 
\beqa
\begin{array}{cccc}
{\bf D3-E3_r}, {\bf E3_r-D3} & {\rm Fermions \ \ } & \sum_{i=0}^{N-1} \,
[\,(n_i,{\ov v^r}_{i-\frac 12 a_r}) + (v^r_i,{\ov n}_{i-\frac 12 a_r}) \, ] &
a_r\; {\rm even} \cr
& & \sum_{i=0}^{N-1} \, [\,(n_i,{\ov v^r}_{i-\frac 12 (a_r+1)}) + (v^r_i,{\ov
n}_{i-\frac 12 (a_r+1)}) \, ] & a_r\; {\rm odd}
\label{E33susy}
\end{array}
\eeqa
where again $r$ denotes the complex plane transverse to the $E3^r$
instanton and $a_r$ is defined in Appendix A.. 
There will be couplings between two such fermionic zero modes 
$\eta^{3-E3_r}$, $\eta^{E3_r-3}$ and $D3-D3$ chiral superfields of the form
(see eq.(\ref{superp}))%
\beqa
\sum_{i=0}^{N-1} 
\Tr ( \Phi^r_{i,i+a_r}\ \  \eta^{3-E3_r}_{i+a_r,i+\frac {a_r}{2}}\ \
  \eta^{E3_r-3}_{i+\frac {a_r}{2},i} ) \ \ .
\label{inssuperpz3}
\eeqa

\subsection{E3-D7 instanton zero modes}

If there are $D7^s$ branes which are passing through the singularity 
the $E3^r$ instantons will necessarily intersect  them and there will
generically 
be further instanton fermionic zero modes. The structure of the intersection and of the possible zero mode depends on the 4-cycles wrapped by the E3- and the D7-branes.
We are going to consider here $r\not=s$ 
where the intersection is on the complex plane transverse to the directions $r,s$ (namely 
$z_r=z_s=0$) 
and we briefly discuss the $r=s$ case below. The worldvolumes are   non-compact, 
hence the actual existence of zero modes depends on boundary 
conditions at infinity. 
Equivalently, if we consider that the singularity is eventually embedded into a CY manifold, 
the instanton and D7-branes have finite volume on the CY, and the E3-D7 spectrum depends on the 
details of the compactification. 
To give a concrete example, one may consider that in a global toroidal embedding of the model there may be Wilson lines on the D7-brane along the relevant 2-torus, such that certain of these E3-D7 zero modes are projected out. Likewise, the E3 instanton may have also Wilson lines which may project out some E3-D7 states. Note that in the case of of $O(1)$ instantons one can still have a discrete $Z_2$ 
Wilson line, which automatically removes all E3-D7 zero modes.

In studying the local model, we consider this not to be the case, and will abuse language by denoting the E3-D7 fields as zero modes. Thus we will need to remove these fields if necessary in concrete examples which involve ingredient projecting out such zero modes (like the Wilson lines mentioned above).

The $E3^r-D7^s$ sector has 8 space-time dimensions with DN boundary conditions.
Thus there are no bosonic zero modes. Since there is a twisted complex plane with NN boundary conditions, the multiplicities and quantum numbers  of the fermion zero modes is analogous to that of $D7^r-D7^s$ systems. One thus gets fermionic zero modes 
given by:
\beqa
\begin{array}{cccc}
{\bf D7^s-E3_r}, {\bf E3_r-D7^s} & {\rm Fermion} & \sum_{i=0}^{N-1} \,
[\,(u^s_i,{\ov v^r}_{i-\frac 12 a_t}) + (v^r_i,{\ov u^s}_{i-\frac 12 a_t}) \, ] &
a_t\; {\rm even} \cr
& & \sum_{i=0}^{N-1} \, [\,(u^s_i,{\ov v^r}_{i-\frac 12 (a_t+1)}) + (v^r_i,{\ov
u^s}_{i-\frac 12 (a_t+1)}) \, ] & a_t\; {\rm odd}
\label{ins7specsusy}
\end{array}
\eeqa
with $t\not=s\not=r\not=t$.
We will denote them $\eta^{rs}$ , $\eta^{sr}$ 
 with two superindices 
indicating they come from the overlap of $D7^s$ and $E3^r$
branes.

These $E3^r-D7^s$ fermionic zero modes have three types 
of couplings to chiral fields on
33 and 37 sectors:

\begin{itemize}

\item
{\bf $(E3^r-D3)(D3-D7^s)(D7^s-E3^r)$}

This is a coupling between a chiral superfield in a $D3-D7^s$ sector
to two fermionic zero modes from $E3^r-D3$ and $D7^s-E3^r$ respectively.
\beqa
 \sum_{r\not=s}
 \eta^r \ \Phi^{(37_s)}  \  \eta^{rs}\ \  
\label{ins11uperpz3}
\eeqa

\item
{\bf $(E3^r-D7^s)(D7^s-D7^s)(D7^s-E3^r)$}

Let us assume that the fields in the D7$^s$-D7$^s$ sector, which propagate on the non-compact $z_s$ 2-plane, have boundary conditions leading to zero modes (we implicitly make this assumption in forthcoming similar analysis). Then there is a coupling between a chiral superfield in a $D7^s-D7^s$ sector
to two fermionic zero modes from $E3^r-D7^s$ and $D7^s-E3^r$ respectively.
\beqa
\epsilon ^{rst}\ 
 \eta^{rs} \ \Phi_t^{(77_s)}  \  \eta^{sr}\ \
\label{ins12uperpz3}
\eeqa

\item
{\bf $(E3^r-D7^s)(D7^s-D7^t)(D7^t-E3^r)$}

This is a variation of the previous one but with $\eta^{rs}$ zero modes
coupling chiral fields in a mixed $D7^s-D7^t$ sector. 
\beqa
 \eta^{rt} \ \Phi^{(7_t7_s)}  \  \eta^{sr}\ \
\label{ins13uperpz3}
\eeqa
with $r\not=s\not=t\not=r$.

Note that in the case of the last two couplings, a vev for the chiral 
fields $\Phi_t^{(77_s)}$ and/or $\Phi^{(7_t7_s)}$ would give mass to
$E3^r-D7^s$ fermion zero modes. In such a case one has to integrate out
appropriately those fermionic zero modes to obtain the correct effective 
instanton action. Such vevs may be triggered in the presence of
non-vanishing FI-terms for $U(1)$'s living on the 
worldvolume of D7-branes (in fact this is equivalent to the appearance of an insertion of the 
corresponding field in the effective 4d superpotential, taking a constant value if the field 
acquires a vev).

Similar analysis can be carried out for $r=s$ for different Chan-Paton actions 
for E3- and D7-branes, namely for E3- and D7-branes which wrap the same 4-cycle, but carry different world-volume gauge bundles. In that case one can check there are again no massless bosonic zero modes but three copies of fermionic zero modes $(E3^r-D7^r)_i$, $i=1,2,3$.
They give rise to couplings to $(D7^r-D7^r)_j$ as well as $(D3-D7^r)$  $D=4$ fields
analogous to those just discussed. We will not discuss them in more detail here.
Finally, as we said, the case of overlapping E3- and D7-branes with 
same Chan-Paton action correspond to  brane instantons with interpretation as  standard gauge
theory instantons.

\section{Action and charges of the instantons}
\label{action}

The action of these instantons has two pieces corresponding to a global piece
which depends on the 4-cycle $\Sigma_4^r$  wrapped by the $E3^r$ (which depends on
untwisted Kahler moduli $T_i$)  and a local piece depending on the
twisted moduli $\phi_k$ at the singularity. Thus the classical action of
the instanton should have the two pieces
\beq
S_{E3^r}\ =\ S_{E3^r}^{unt}\ +\ S_{E3^r}^{twist}\ = \ T_{\Sigma_4}^r \ +\ 
\sum_{k=0}^{N-1}\ d^r_k\ \phi_k
\label{actins}
\eeq
Here $T^r$ will be some linear combination of untwisted Kahler moduli
characteristic of the cycle of the $E3^r$ in the bulk. Its real part is 
controlled by the volume of the wrapped toroidal 4-cycle. The 
$\phi_k$ are closed string twisted moduli associated to the singularity.
In the orbifold configuration, the background values of the twisted moduli are 
zero, $\phi_k=0$, so the instanton amplitude is controlled by $ReT^r$. The 
coefficients $d^r_k$ will be computed below.
As we have just seen, there are additional pieces in the action coming
from couplings among  bifundamental 33 chiral fields $\Phi_r^{\alpha \beta }$
and the $E3^r-D3$ zero modes $\eta^\alpha_i$, $\eta^\beta_j$
of the general form
\beq
S_{E3^r}' \ =\ \sum_{i,j,r}\ c_{ij}^r\ \eta^{\alpha}_i \Phi_r^{\alpha \beta }\eta^\beta_j
\eeq
Integration over the fermionic zero modes  
gives rise to a non-perturbative superpotential
\beq
e^{-S_{E3^r}}\ \int [d\eta^\alpha ][d\eta^\beta]
\ e^{-\sum_{i,j,r}\ c_{ij}^r\ \eta^{\alpha}_i \Phi_r^{\alpha \beta }\eta^\beta_j}
\ \propto \ e^{-S_{E3^r}}\ det(\Phi_r) \ \ .
\label{inssuper}
\eeq
Let us now check that  this induced operator is invariant under
the gauged $U(1)$ symmetries on the 4d spacefilling D3-branes (Note that these 
arguments are valid even if the instanton has additional zero modes, and thus 
leads to a higher F-term in the 4d effective action) \footnote{This kind of computation is 
a particular case of the general argument in the appendix of \cite{iu}, 
which applied to general systems of D-branes and instantons. 
From this viewpoint, the discussion below amounts to a careful computation 
of the couplings of the branes to the RR fields in the orbifold CFT.}. 

As shown in \cite{iru} it 
is only the twisted moduli which 
are shifted under the $U(1)$ gauge symmetries living on the D3-,D7-branes. In 
particular consider a stack of  $D3$-branes living at a
 $\IR^6/\IZ_N$ singularity. Consider the $U(1)_a$ group associated to
one of the $U(n_a)$ factors with CP matrix $\lambda _a$. Following the rules 
in \cite{dm}, it was shown
in \cite{iru} that there are couplings
\beq
\sum_{k=0}^{N-1}\ \sqrt{c_k}\ Tr(\  \gamma_{\theta^k}^{D3}\ \lambda_a )
\times ( F_a \wedge B_k) 
\eeq
where $B_k$ are RR 2-forms in the $k^{th}$ twisted sector, thus associated to 
the 
singularity. The twist CP matrix $\gamma_{\theta ^k}^{D3}$ is defined in Appendix A. 
The $B_k$  are Poincare dual to scalars $b_k=Im\phi_k$ which then transform under
a $U(1)_a$ gauge transformation of parameter $\Lambda(x)_a$ like 
\beq
b_k\ \longrightarrow \ b_k \ +\ \sqrt{c_k} \  Tr(\  \gamma_{\theta^k}^{D3}\ \lambda_a )
\ \Lambda(x)_a 
\eeq
The value of the $c_k$ coefficients is given below. 
Now, the twisted piece of the action of the $E3^r$ instanton may be obtained 
from the corresponding DBI+CS action. In particular the coupling to the 
imaginary part of the twisted field is topological, and 
may be inferred from the known couplings of the $F\wedge F$ term 
of a $D7^r$ brane to the twisted RR fields $b_k$. 
This is because a 4d  $F\wedge F$ background on a D7$^r$-brane must carry the 
same topological couplings as an E3$^r$-brane. Notice however that the D7-brane 
is in general not present in the configuration, and we merely use it as a trick 
to obtain the couplings, which can be computed by other techniques. The 
couplings for the D7-branes were also found in \cite{iru} to be given by 
\beq
\frac {1}{N}
\sum_{k=0}^{N-1}\ \sqrt{c_k}\ Tr(\  (\gamma_{\theta^k}^{D7^r})^{-1} \ \lambda_i^2 )
\ b_k \ \times ( F_i \wedge F_i)
\eeq
Including also the real part of $\phi_k$, mentioned above,  one then has for
 the  $E3^r$ instanton action    
\beq
S_{E3^r} \ = \ T_{\Sigma_4}^r \ +\ \frac {1}{N} \  
\sum_{k=0}^{N-1}\ \sqrt{c_k}\ Tr(\  (\gamma_{\theta^k}^{E3^r})^{-1} \ 
(\lambda_b)^2 )
\ \phi_k  
\label{actinss}
\eeq
with $\lambda_b$ a  CP matrix  of the $E3^r$.
One can then compute how the instanton action transforms under
a $U(1)_a$ symmetry. 
For a $E3^r$ instanton transverse to the $r$-th complex plane one has 
\cite{iru}
\beq
c_k \ =\ 2\ sin(\pi ka_r/N) \ =\ -i\  (\ \alpha^{\frac {ka_r}{2}} \ -\ 
\alpha^{- \frac {ka_r}{2}})
\eeq
with $\alpha=exp(i2\pi/N)$. The coefficients $\sqrt{c_k}$ of the coupling of the D-branes to 
the RR fields can be obtained by factorization of a cylinder diagram \cite{abd}. 

With the CP $\gamma_{\theta^k}$ matrices defined
above one then obtains:
\beqa
S_{E3^r} \ \longrightarrow & \ &
S_{E3^r} \ +\ 
i\frac {n_av_b}{N} \ \sum_{k=0}^{N-1}\ c_k\ \alpha^{ak}\ \alpha^{-bk}\ 
\Lambda(x)_a \\
& =& S_{E3^r} \ +\
\frac {n_av_b}{N} \ \sum_{k=0}^{N-1}\ (\  \alpha^{k(\frac {a_r}{2}+a-b)}
\ -\ \alpha^{k(-\frac {a_r}{2}+a-b)}\ )\ \Lambda(x)_a \\
 \ & =& \ S_{E3^r} \ +\ n_av_b\ 
(\delta_{b,a+{\frac {a_r}{2} }}
\ -\ \delta_{b,a- {\frac {a_r}{2}}})\ \Lambda(x)_a
\label{trans3}
\eeqa
Given the above noted relation between E3$^r$-branes and 
D7$^r$-brane instantons, ie 4d $F\wedge F$ backgrounds on D7-branes, this 
computation is quite analogous to the way the  Green-Schwarz  mixed anomaly 
cancellation takes place between a $U(1)$ from $D3$ branes and gauge groups 
from $D7$'s. Therefore, the appearance of the Kronecker deltas, which count the 
number of multiplets in the mixed D7/E3-D3 sector, is not a surprise.

One can now easily check that the charge transformation obtained is just 
opposite  to the total  $U(1)_a$ charge of $E3^r-D3$ fermionic zero modes 
transforming like (we are taking $a_r$ even):
\beq
(n_{b+\frac {a_r}{2}} , {\bar v}_b)\ +\ (v_b, {\bar n}_{b-\frac {a_r}{2}}  ) \ \ .
\eeq
This implies that the complete instanton effective vertex (\ref{inssuper}), 
which includes the exponential term and the insertions of 4d fields due to the 
integration of charged fermion zero modes, is indeed gauge invariant.

\medskip

If the $E3^r$ instanton intersects some $D7^s$ brane, in general 
its action will also transform under $U(1)$'s living on the $D7^s$ branes.
In general a $U(1)_c$ gauge symmetry inside a $D7^s$ brane will
couple to twisted moduli like 
\beq
\sum_{k=0}^{N-1}\ \sqrt{c_k}\ Tr(\  \gamma_{\theta^k}^{D7^s}\ \lambda_c )
\times ( F_c \wedge B_k)
\eeq
We are considering $s\not=r$ (for $r=s$ a similar discussion can be carried out) and now the 
$c_k$ factor will be given by
$c_k \ =\ 2\ sin(\pi ka_t/N)$ with $t\not=r\not=s\not=t$. Thus $z_t$ is the 
complex direction with NN boundary conditions in the $E3^r-D7^s$ system.
Now the RR twisted fields transform with respect to a $D7^s$ $U(1)_c$ like
\beq
b_k\ \longrightarrow \ b_k \ +\ \sqrt{c_k} \  Tr(\  \gamma_{\theta^k}^{D7^s}\ \lambda_c )
\ \Lambda(x)_c
\eeq
and $c_k$ is as indicated above. One can again repeat the analysis
we made for $D3$-branes  
and find that under a $U(1)_c$ gauge transformation the action transforms 
like
\beq
S_{E3^r}\ \longrightarrow
 \ S_{E3^r} \ +\ u_cv_b\
(\delta_{b,c+{\frac {a_t}{2} }}
\ -\ \delta_{b,c- {\frac {a_t}{2}}})\ \Lambda(x)_c  \ \ .
\label{trans7}
\eeq
One can then check that this transformation corresponds to the opposite of the  
overall $U(1)_c$ charge of $E3^r-D7^s$ fermionic zero modes transforming like:
\beq
(u_{c+\frac {a_t}{2}} , {\bar v}_c)\ +\ (v_c, {\bar u}_{c-\frac {a_t}{2}}  )
\eeq
which we already discussed are indeed present, see (\ref{ins7specsusy}). Thus again one recovers
a fully gauge invariant operator. In the presence of $D7^s$ branes
though the induced superpotentials will involve 
chiral fields from all $33$, $37$ and $77$ sectors, as we will see
in the specific examples.

Note that the superpotential will in general be induced only
if the number of universal neutral fermion zero modes of the $E3^r$ instanton
is two, providing us for the superspace measure. This is guaranteed if there
is in addition an orientifold projection and the instanton has 
an $O(1)$ CP factor. In this connection note that
all the above expressions were obtained for the case of branes at
orbifold (not orientifold) singularities. The same expressions may be used for the 
orientifold case simply recalling the mapping between branes with 
$\alpha^k$ and $\alpha ^{N-k}$ CP factors
upon the orientifold action.

\section{ Applications}
\label{examples}

\subsection{A local $SU(3)_c\times SU(3)_L\times SU(3)_R$ model from $D3$-branes at a 
$\IZ_3$ singularity }

Consider the simplest case, in which no $D7^s$ branes are present and we just have a
stack of $D3$ branes at a singularity. Then twisted RR tadpole conditions 
dictate (if the $D3$'s are away from orientifold planes) that 
$Tr\gamma_{\theta^k,3}=0$ (for all twists $\theta^k$ with the origin as only fixed point). The 
gauge group has then the structure
$\Pi_{i=0}^{N-1}U(n)$ with $Nn$ the total number of $D3$'s.
A phenomenologically interesting example is the case with 
$N=n=3$ in which we sit at a $Z_3$ singularity and the gauge group
is $U(3)_c\times U(3)_L\times U(3)_R$. This structure contains the
SM gauge group and it has been termed 'trinification' in the unified
model-building literature
\cite{trinification}. There are three generations with chiral
multiplets from the 33 sector transforming like 
\beq
3(3,{\bar 3},1)\ +\ 3({\bar 3},1,3)\ +\ 3(1,3,{\bar 3})
\eeq
These three multiplets (in 3 copies) contain respectively the
left-handed quarks, right-handed quarks and the lepton/Higgs fields 
(plus additional vectorlike leptons). Now, the perturbative superpotential 
is given by 
\beq
W \  = \  \sum_{r,s,t=1}^3\; \epsilon_{rst} \; 
(3,{\bar 3},1)^r({\bar 3},1,3)^s(1,3,{\bar 3})^t \ \ .
\eeq
Note that, with the Higgs multiplets inside $(1,3,{\bar 3})$, 
 this includes some Yukawa couplings for the 
quarks. However there are no lepton Yukawa couplings
\footnote{The absence of some perturbative either quark or lepton Yukawa
couplings is a quite general property in semirealistic 
models of branes at singularities. See the left-right symmetric example in Section 
\ref{leftright}.} since they would require the presence of $(1,3,{\bar 3})^3$
couplings, which are  forbidden by the $U(1)_L\times U(1)_R$
gauge symmetry. 

Let us assume  that there is a $E3^r$ which has $O(1)$ CP symmetry and
goes through this $Z_3$  singularity. It will have its CP factor =1.
The $D3$ twist CP matrix will be
\beqa
\gamma_{\theta,3}& = &\diag(\id_3,\alpha \id_3, \alpha^2 \id_3)
\eeqa
Take for definiteness $a_r=-2$ then there are $E3^r-D3$ zero modes
transforming like 
\beq
\eta^r\ =\ (1,1,3) \ \ ;\ \ {\bar \eta}^r\ =\ (1,{\bar 3},1)
\eeq
which have couplings to the r-th lepton chiral field
\beq
\eta^r\ (1,3,{\bar 3})^r\ {\bar \eta}^r \ .
\eeq
Upon integration of these charged zero modes a superpotential coupling
\beq
W_{leptons}^r\ =\ e^{-S_{E3^r}}\ 
\epsilon_{abc}\epsilon_{def}\ (1,3^a,{\bar 3}^d)^r
(1,3^b,{\bar 3}^e)^r(1,3^c,{\bar 3}^f)^r
\eeq
is obtained for the r-th generation of leptons. 
From eq.(\ref{trans3}) one obtains
\beq
S_{E3^r}\ \longrightarrow \ S_{E3^r}\ +\ 3\Lambda_{U(1)_L}\ -\ 3\Lambda_{U(1)_R}
\eeq
so that indeed the operator is fully gauge invariant.
Instantons $E3^s$ transverse to the other two complex planes 
would give rise to leptonic Yukawa couplings for the 
other two generations.
This is a simple example of how this class of instantons
may give rise to superpotential couplings of 
phenomenological interest. Note that these couplings are presumably suppressed with respect to 
the quark one, but need not be negligibly small. 
Indeed, the SM gauge couplings in this 
model are given by the inverse of the real part of the 4-dimensional dilaton field $S$
(plus a twisted moduli piece analogous to the second term
in eq.(\ref{actinss}) ). These are totally independent from  the
instanton action (\ref{actinss}) which is rather controlled
by a  combination of untwisted Kahler moduli $T^r$ which may be
relatively small without any phenomenological constraint 
dictated by the observed values of the SM couplings.

\subsection{ A global tadpole free GUT-like  example}

Our previous example was a local model from $D3$-branes 
at  a singularity.
We would like now to show in a simple example how the
generation of open string superpotentials can take place
in simple, globally consistent (RR tadpole free) 
compact models. We will take the simplest compact
Type IIB 4-D orientifold  which one can build, a 
$Z_3$ orientifold with O3-planes \cite{abpss,lpt} (for a discussion of D-brane instantons in the $\IZ_3$ 
orientifold, with a different distribution of D-branes, see \cite{bk}). 
Similar
effects should be found at more complicated toroidal orientifolds.

Let us consider type IIB on the $T^6/\IZ_3$ orbifold, modded out by the orientifold action
$\Omega (-1)^{F_L}\ R_1\  R_2\ R_3$, with $R_i$ being a reflection on the
$i^{th}$ plane. There are 64 orientifold three-planes (O3-planes), which
are localized at points in the internal space. To cancel their untwisted RR charge
we need a total of 32 D3 branes. There are also  27
orbifold fixed points which may be labeled by integers $(m,n,p)$, $m,n,p=0,\pm 1$. 
 Among these 27 points, only the
origin $(0,0,0)$ is fixed under the orientifold action, hence it is an
orientifold singularity.  The
cancellation of tadpoles at this point requires
\beq
3\ \Tr\gamma_{\theta,3} + (\Tr\gamma_{\theta,7}
- \Tr\gamma_{\theta,{\bar{7}}}) = -12
\label{tadpoin}
\eeq
In our case with no $D7$-branes present the condition is $\Tr\gamma_{\theta,3}=-4$.
An $SU(6)$ GUT model may be constructed in the following way.
We can locate 14 $D3$-branes at the orientifold plane at  the origin with 
CP twist matrix
\beqa
\gamma_{\theta,3}& = &\diag(\alpha \id_6, \alpha^2 \id_6, \id_2)\ 
\eeqa
and the remaining 18 $D3$ e.g. in the bulk (e.g. in 3 orbifold/orientifold invariant
groups of 6 $D3$ branes), away from the origin in any of the tori.
The orientifold operation exchanges D3-branes with CP factors $\alpha$ and $\alpha^2$.
The gauge group is $U(6)\times O(2)$ with chiral fermion content
\beq
3({\overline {15}},0)\ +\ 3(6, +1)\ +\ 3(6,-1)\ .
\eeq
These representations decompose as ${\overline {15}}={\overline {10}}+{\overline {5}}$
and $6=5+1$ under the $SU(5)$ subgroup of $SU(6)$, so the model contains three standard $SU(5)$ 
generations ${\overline {10}}+5$
and three sets of $5+{\overline 5}$ Higgs fields. With the usual SM embedding in $SU(5)$, the
model has lepton and  D-quark Yukawa couplings from quiver couplings of the
form $(\overline {15},0)(6,1)(6,-1)$. However U-quark Yukawa couplings would be contained
in   ${\overline {15}}{\overline {15}}{\overline {15}}$ couplings which are perturbatively
forbidden by the $U(1)$ symmetry.

Now these  compact orientifolds admit BPS euclidean branes $E3^r$ which wrap
two 2-tori and are transverse to the r-th torus. If they sit at the
origin the projection will be such that,  if  $\gamma_{\theta,E3^r}=1$, the   CP symmetry
will be $O(1)$ and hence the number of neutral instanton zero modes
will be adequate to create a superpotential
\footnote{These $E3^r$ instantons will in general
contain in their worldvolume other orbifold fixed points without D3-branes.
This means that the instanton action will also contain pieces
involving these other twisted fields. However the latter are inert
under the $U(1)$ transformations of the D3-branes at the orientifold
plane and hence those extra pieces do not play
any role in our discussion.} . There is one multiplet
of charged fermion zero modes for each $E3^r$
\beq
\eta^r_a\ =\ (6_a,0)^r
\eeq
coupling to the antisymmetric chiral fields $\Phi^r={\overline {15}}^r$  like
\beq
\eta^r_{a}\eta^r_b\Phi^r_{ab} \ \ .
\eeq
Upon integrating over zero modes a cubic superpotential  is generated
\beq
W_6\ =\ \sum_r\ e^{-S_{E3^r}}\ \epsilon^{abcdef}\Phi^r_{ab} \Phi^r_{cd}\Phi^r_{ef}
\eeq
which includes the U-quark Yukawas which were perturbatively absent.
The instanton action transforms like 
\beq
S_{E3^r}\ \longrightarrow \ S_{E3^r}\  -\ 6\Lambda_{U(1)_3}
\eeq
so that again the operator is fully gauge invariant.

Note that the size of these Yukawas  will depend on the corresponding action
which has the form previously discussed (\ref{actinss}) with now
the $T^r$ being actually the untwisted Kahler moduli of this
orientifold. Thus the actual values of the couplings  is sensitive to
the overall sizes of the different dimensions.

This GUT model is not fully realistic since as it stands it lacks the
required Higgs multiplets to do the breaking down to the SM. Still it
exemplifies in a global tadpole free model how instantons may give 
rise to phenomenologically interesting couplings.
In the context of the compact $Z_3$ orientifold
the generation of such terms was recently pointed out in
\cite{bfm}. Instanton induced Yukawa couplings in an $SU(5)$ model
from a local intersecting $D6$-brane configuration were  also considered recently
in    \cite{bclrw}. 

Other globally consistent compact orientifold models are expected to
present the same type of instanton induced couplings. As an 
additional example we discuss the case of the $\IZ_7$ orientifold
model in an appendix.

\subsection{An example with multi-instanton superpotential}

Let us consider  the same $\IZ_3$ toroidal orientifold, 
but with a different distribution of D3-branes. 
It illustrates a phenomenon in \cite{geu}, in which 
instantons with additional fermion zero modes 
(beyond the two required Goldstinos) can still contribute to the superpotential, 
if the extra zero modes are lifted (or soaked up) by another instanton. 
We dub this phenomenon "instanton symbiosis".

Consider the same $T^6/\IZ_3$ orientifold as above, and introduce a set of D3-branes at the origin, with Chan-Paton orbifold action $\gamma_{\theta}=\diag (e^{2\pi i/3}\id_4,e^{4\pi i/3}\id_4)$. This leads to a $U(4)$ gauge theory with three chiral multiplets in the two-index antisymmetric representation, and no superpotential. The remaining 24 D3-branes can be located as four stacks of bulk D3-branes, and are irrelevant to our discussion.

Let us focus on the possible non-perturbative effects for the local configuration at the $\IZ_3$ orientifold singularity a the origin.
There is an $O(1)$ instanton, corresponding to an euclidean D-brane 
filling the unoccupied node in the quiver (that is a fractional E$(-1)$-brane 
with Chan-Paton orbifold action $\gamma_{\theta,E(-1)}$. It has two neutral fermion 
zero modes, and charged fermion zero modes (in 3 copies)  in the fundamental of $U(4)$, 
with cubic couplings to the 4d chiral multiplets in the ${\bf 6}$. 
It gives rise to a non-perturbative superpotential of very high order 
in the 4d chiral multiplets, which will not interest us.
In addition, there is a $U(1)$ instanton, arising from E$(-1)$-branes 
with Chan-Paton action $\gamma_{\theta,E(-1)}=\diag(e^{2\pi i/3}, e^{4\pi i/3})$.
 It corresponds to a gauge theory instanton, so its effects are more
 suitably analyzed using field theory arguments.
Forgetting about the $U(1)$, which is massive by $BF$ couplings,
 the 4d gauge theory can be regarded (since $SO(6)\simeq SU(4)$)as an $SO(N_c)$
  theory with $N_f=N_c-3$ flavors in the vector representation \cite{is},
 for $N_c=6$, hence $N_f=3$. The case $N_f=N_c-3$ is somewhat analogous to 
the case $N_f=N_c+1$ for $SU(N_c)$ SQCD. In particular, instantons have extra zero modes 
beyond the two Goldstinos, which are not lifted. These instantons do not generate a superpotential term, but rather induce higher order F-terms, as in \cite{Beasley:2004ys}.

In addition there may be effects from $O(1)$ instantons described by euclidean 
E3$_r$-branes. These are similar to those considered in the previous Section, and lead to non-perturbative mass terms $m_r \sim M_s\,  e^{-T_r}$ for the chiral multiplets in the ${\bf 6}$.
This model and the above discussion have already appeared in \cite{bk}.

In the following we would like to have a closer look at the effects of the E3$_r$, and argue that they have a non-trivial effect on the gauge instantons and implement the non-perturbative lifting of zero modes advocated in \cite{geu}. This example thus illustrates  another interesting application of E3$_r$-brane instantons, and shows that the effects in \cite{geu} naturally arise already in familiar toroidal orientifold models.

From the spacetime viewpoint, the mass terms induced by the 
E3-brane instantons have a non-trivial effect on the dynamics of the $SU(4)$ gauge theory. Indeed in the far infrared one can integrate out the massive flavours and be left with a pure $SU(4)$ theory, which has the familiar non-perturbative gaugino condensate superpotential
\beqa
W\, \sim\, \Lambda'^3
\label{puresymsupo}
\eeqa
where the pure SYM scale $\Lambda'$ in the IR is related to the UV scale 
$\Lambda$ by matching of scales $\Lambda^{9}\prod_r m_r = \Lambda'^{12}$. 
In our case the UV scale is 
$\Lambda=M_s 
e^{-1/(9 g_YM^2)}=M_s e^{-S/9}$, where the factor of $9$ arises from the beta function (proportional to $3(N_c-2)-N_f$), and where $S$ is the modulus giving the gauge coupling of the $U(4)$ theory at high scales (essentially the 4d dilaton, plus corrections from twisted moduli). Hence we have
\beqa
W\, \sim \, ( \, e^{-S}\, \prod_r \, e^{-T_r}\, )^{\frac 14}\, M_s^3 
\label{puresym2}
\eeqa

The above argument shows that there is a non-trivial effect of the E3$_r$ instantons on the E$(-1)$-brane gauge instantons, so that the latter can induce a superpotential. 
In fact, this is a particular case of the non-perturbative lifting of fermion zero modes in \cite{geu}, where the E3-brane instantons induce a lifting of the additional fermion zero modes of the E$(-1)$-brane instantons. The overall process corresponds to a multi-instanton process in spacetime, where all fermionic external legs of the E$(-1)$-brane instanton vertex, except two, are soaked up by the E3$_r$-brane instanton vertex. The interpretation as a multi-instanton process agrees nicely with the exponential dependence of (\ref{puresym2}) on the moduli.

Let us sketch a simplified version of the mechanism. In the simultaneous presence of the instantons, there are fermion zero modes $\lambda_r,{\tilde \lambda}_r$ in the E3$_r$-E(-1) sector. These couple to different pairs of the extra fermion zero modes in the E$(-1)$-D3 sectors, which we denote $\chi_r$, ${\tilde \chi}_r$ via a quartic interaction
\beqa
S_{\rm inst}\, =\, \sum_r \lambda_r {\tilde \lambda}_r \chi {\tilde \chi}_r
\eeqa
Notice that the $r$ subindex for the $\lambda$'s denotes the open string sector, while for the $\chi$'s it simply denotes the set of $\lambda$'s to which they couple. The above argument is somewhat sloppy, since there are no microscopic quartic interactions in orbifolds. However they can be regarded as effective interactions upon integrating over bosonic zero modes from open strings between the different instantons (see \cite{geu} for details). We stick to this simplified discussion.

Upon integrating over the fermion zero modes $\lambda_r$, ${\tilde \lambda}_r$ of the E3$_r$-brane instantons, we find they induce an effective interaction on the world-volume action of the E(-1)$_r$-instanton.
\beqa
\delta S_{{\rm E}(-1)}\, =\, \sum_r \, e^{-T_r} \, \chi_r {\tilde \chi}_r
\eeqa
So the additional fermion zero modes of the E$(-1)$-brane instanton are lifted. 
In other words, one can pull down insertions of  $\delta S_{\rm E(-1)}$ to soak 
up the integrations over the fermionic collective variables $\chi$, ${\tilde 
\chi}$. Integrating over the latter, the E$(-1)$-brane instanton thus leads to 
a 4d non-perturbative superpotential term precisely of the form 
(\ref{puresym2}). The power of $1/4$ arises from the fact that we are dealing 
with fractional instantons \footnote{In more precise terms, the standard gauge 
theory instanton contributes to a 4d correlator involving four pairs for 
fermions, out of which one can extract two-fermion correlators by clustering, 
as is familiar in the description of the gaugino condensate superpotential in 
$N=1$ pure SYM.}. One can check  that the complete superpotential is invariant
under transformations of the $U(1)$ symmetry in $U(4)$. This is correlated
with the fact that the complete 4-instanton system has zero intersection
number (namely zero net number of chiral fermion zero modes) with the
D3-branes.

A similar discussion would apply to other examples where euclidean brane 
instantons have been argued to modify the infrared dynamics of gauge theory 
sectors e.g. \cite{fkgs,abfk}.

\subsection{Instanton Induced SUSY Breaking}

The same $Z_3$ orientifold considered above 
may be used to construct a compact model in
which SUSY-breaking a la Fayet may be implemented along the lines
of the recent work \cite{aks}. 
In the present case we locate
 8 $D3$-branes with CP twist matrix
\beqa
\gamma_{\theta,3}& = &\diag(\alpha \id_4, \alpha^2 \id_4)
\eeqa
at the orientifold point at the origin. 
This is enough to cancel twisted tadpoles.
We then locate the remaining
24 $D3$-branes e.g. in the bulk (in 4 orbifold/orientifold invariant
groups of 6 $D3$ branes), away from the origin in any of the tori.
The gauge group from the branes at the origin is $U(4)$ with 
three chiral fields $\Phi^r$  ($r=1,2,3$) transforming like $(6,-2)^r$
under $SU(4)\times U(1)$. 
There is one multiplet
of charged fermion zero modes for each $E3^r$
\beq
\eta^r\ =\ (4,1)^r
\eeq
coupling to the chiral fields like 
\beq
\epsilon^{abcd}\eta^r_{a}\eta^r_b\Phi^r_{cd} \ \ .
\eeq
Upon integrating over zero modes a superpotential is generated
\beq
W_6\ =\ \sum_r\ e^{-S_{E3^r}}\ \epsilon^{abcd}\Phi^r_{ab} \Phi^r_{cd}
\eeq
which gives mass terms to the three 6-plet  matter fields.
These are the mass terms which were mentioned in the previous
section.
Again,  the actual values of the masses is sensitive to
the overall sizes of the different dimensions.
Now,  there is also a $U(1)$ D-term potential of the form
\beq
V_{U(1)}\ =\ \frac {1}{\lambda}\ (\sum_r\ -2|\Phi^r|^2\ +\ \xi)^2
\eeq
where $\xi$ is a (field dependent) FI-term \footnote{ In fact, supersymmetry relates this to 
the $BF$ coupling, discussed above, from which such FI-term is given by
$
\xi \ =\frac {1}{2}\ 
\sum_{k=0}^{N-1}\ \sqrt{c_k}\ Tr(\  (\gamma_{\theta^k}^{D3}
- \gamma_{\theta^{N-k}}^{D3})
\ \lambda_a )\phi_k.$}.
Assume that there is some mechanism stabilizing all kahler moduli
(twisted and untwisted), so that $\xi$ will be a fixed parameter. The structure is now that of 
a variation of the Fayet-Iliopoulos mechanism for SUSY-breaking. Indeed, minimization of 
the D-term requires some of the 6-plets $\Phi^r$ to get a vev. But such a vev
would give rise to a non-vanishing vacuum energy, due to the 
instanton-induced mass terms. Hence SUSY is spontaneously broken a la 
Fayet. The scale of SUSY breaking is of order:
\beq
F_{\Phi^{r}}\ \simeq \ e^{-S_{E3^r}}\ \sqrt{\xi}
\eeq
with $\Phi^r$ the scalar with the largest instanton suppression 
$e^{-S_{E3^r}}$.
Note that this could be used as  the SUSY-breaking sector of 
a trinification model like that in the previous subsection, with gauge group
$U(3)\times U(3)\times U(3)$ and three generations.

The present  example differs from \cite{aks} in several ways.
The latter considered a $\IZ_3$ orbifold of the conifold singularity, where the
effects come from  euclidean fractional $E1$ instantons, rather than  Euclidean $E3$ 
instantons. Our example is also a global tadpole-free model, rather than a local one.

\subsection{Yukawas and $\mu$-terms in a  Left-Right symmetric model}
\label{leftright}

As we already mentioned, a typical drawback of semirealistic constructions
from $D$-branes at singularities is that some Yukawa couplings are perturbatively 
absent. We would like to show that E3-brane instantons can cure this pathology, and provide an 
specific example in a semirealistic model. Here we consider a semirealistic left-right 
symmetric local $Z_3$ model. This is possibly the phenomenologically most attractive model among the ones constructed in \cite{aiqu}
\footnote{ It is easy to show that E3-brane instantons 
in the alternative Standard Model-like 
examples also considered  
in \cite{aiqu} lead to an unpaired set of D3-E3 fermion zero modes, 
which therefore lead to a vanishing amplitude.
 This unpairing can be regarded as a global $O(1)$ anomaly 
on the instanton world-volume, and signals the presence of some 
uncanceled charge in the non-compact model. In a global
 compactification this charge should cancel, which implies 
(regarding the instanton as a probe of discrete tadpoles, as in \cite{Uranga:2000xp}) 
that these instantons necessarily have extra charged fermion zero modes from 
other sources, and thus necessarily lead to extra unwanted 
insertions in the 4d amplitude, thus rendering this case less interesting.}. Let us review it here (see 
\cite{aiq} for further discussion).

We consider the D3-brane Chan-Paton embedding
\beqa
\gamma_{\theta,3}& = &\diag(\id_3,\alpha \id_2, \alpha^2 \id_2)
\eeqa
The corresponding tadpoles can be canceled for instance by D7$_r$-branes,
$r=1,2,3$ with the symmetric choice $u_0^r=0$, $u_1^r=u_2^r=1$. The gauge
group on D3-branes is $U(3)\times U(2)_L\times U(2)_R$, while each set of
D7$_r$-branes contains $U(1)^2$. The combination
\beq
Q_{B{\rm -}L}= - 2 \left(\frac 13\ Q_{3}\ +\frac 12  \ Q_{L}\ +
\frac12  \ Q_{R}\right)
\eeq
is non-anomalous, and in fact behaves as $\BL$. The spectrum for this
model, with the relevant $U(1)$ quantum numbers is given in
table~\ref{tabpslr}.
\begin{table}[htb] \footnotesize
\renewcommand{\arraystretch}{1.25}
\begin{center}
\begin{tabular}{|c|c|c|c|c|c|c|}
\hline Matter fields  &  $Q_3$  & $Q_L $ & $Q_R $ & $Q_{U_1^i}$ &
   $Q_{U_2^i}$
& $B-L$   \\
\hline\hline {{\bf 33} sector} &  & & & & & \\
\hline $3(3,2,1)$ & 1  & -1 & 0 & 0 & 0 & 1/3  \\
\hline $3(\bar 3,1,2)$ & -1  & 0  & 1 & 0 & 0 & -1/3 \\
\hline $3(1,2,2)$ & 0  & 1  & -1 & 0 & 0 & 0  \\
\hline\hline {{\bf 37$_r$} sector} & & & & & & \\
\hline $(3,1,1)$ & 1 & 0 & 0 & -1  & 0 & -2/3 \\
\hline $(\bar 3,1,1)$ & -1 & 0 & 0 & 0 & 1  & 2/3 \\
\hline $(1,2,1)$ & 0 & 1 & 0 & 0 & -1 & -1 \\
\hline $(1,1,2)$ & 0 & 0 & -1 & 1 & 0 & 1 \\
\hline\hline {\bf 7$_r$7$_r$} sector & & & & & &  \\
\hline $3(1)'$ & 0 & 0 & 0 & 1 & -1 & 0 \\
\hline \end{tabular}
\end{center} \caption{Spectrum of $SU(3)\times SU(2)_L\times SU(2)_R$
model. We present the quantum numbers  under the $U(1)^9$ groups. The
first three $U(1)$'s arise from the D3-brane sector. The next two come
from the D7$_r$-brane sectors, and are written as a single column with the
understanding that {\bf 37}$_r$ fields are charged under $U(1)$ factors in
the ${\bf 7_r7_r}$ sector.
\label{tabpslr} }
\end{table}
We can see that the color triplets from the 37$_r$ sectors can become massive
after the singlets of the ${\bf 7_r7_r}$ sector acquire a nonvanishing
vev, leaving a light spectrum really close to left-right theories
considered in phenomenological model-building, with no chiral exotics.

Although it is possible to embed this configuration in globally consistent compact models
(see section 4.3 in \cite{aiqu}), we 
will restrict to considering the local model. We will consider E3$^r$-brane instantons, wrapping 
non-compact directions, and implicitly assume that they are fixed under some orientifold 
projection in such a way 
that they have an $O(1)$ CP symmetry (see  figure (\ref{instanton5})).
Recall that we consider that the orientifold action does not fix the orbifold point, but 
relates it (and the D3/D7-brane system) to some mirror image $\IZ_3$ singularity.
The $E3$ will go both through the SM $D3$-branes and their mirrors.
%
%
\begin{figure}
\epsfysize=6cm
\begin{center}
\leavevmode
\epsffile{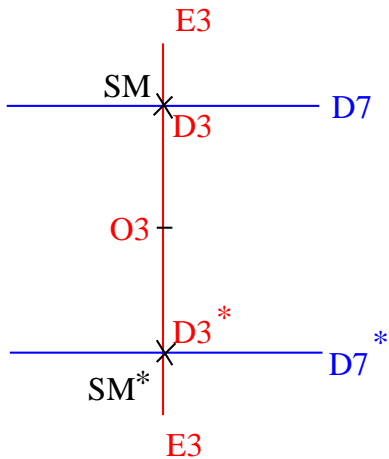}
\end{center}
\caption{Pictorial view of the geometry of the $D3$,$D7$ semirealistic orientifold 
model discussed in the text. The instanton $E3^r$ pass through both the $D3$-branes 
giving rise to the SM group and an  orientifold $O3$ plane. The $D3·$ and $D7$ 
branes have orientifold mirror branes $D3^*$, $D7^*$ whereas the $E3$ instanton
is orientifold invariant.}
\label{instanton5}
\end{figure}

The structure of charged zero modes in this model is as follows:

\end{itemize}
\begin{itemize}
\item
{\bf $E3^r$-$D3$ sector}
\beq
\eta_L^r\ =\ (1,{\bar 2},1)^r_{(-1,0;0,0)} \ \ ;\ \
\eta_R^r\ =\ (1,1,2)^r_{(0,1;0,0)} 
\eeq
where the subindices show the charges under 
$U(1)_L\times U(1)_R\times U(1)_1^r\times U(1)_2^r$.
Here $U(1)_{1,2}$ are the $U(1)$'s associated to each of the 
$D7^r$ branes which are present.
\item
{\bf $E3^r$-$D7^s$ sector}
\beq
\eta_1^{rs}\ =\ (1,1,1)^{rs}_{(0,0;-1,0)} \ \ ;\ \
\eta_2^{rs}\ =\ (1,1,1)^{rs}_{(0,0;0,1)}
\eeq
where here $r\not=s$.

\end{itemize}
As mentioned above, some or all these extra zero modes could be
absent depending on the boundary conditions of the $D7^s$ branes at
infinity. For instance, they are automatically absent for instantons 
with a turned on discrete $O(1)\equiv \IZ_2$ Wilson line (a 
possibility easily implemented in toroidal orientifold models). 
We nevertheless keep them for the moment, with the understanding 
that the corresponding insertions could  be absent for instantons 
with a built-in mechanism to remove E3-D7 zero modes.

Note that  for fixed $r$, since we have  $s\not=r$, there are altogether 
2+2+2(1+1)=8 zero modes. Now the couplings of these  zero modes to the 
chiral fields is as follows:
\beqa
S_{charged}^r\ &=&\ 
\eta_L^r(1,2,{\bar 2})^r\eta_R^r + \\
&+& \ \sum_{s\not=r}\ \eta_L^r(1,2,1)^s\eta^{rs}_2\ +\ 
\sum_{s\not=r}\ \eta_R^r(1,1,{\bar 2})^s\eta^{rs}_1\ + \\
&+& \ \epsilon_{rst}\ \eta^{rs}_1 \Phi^{7_s7_s}_t\eta^{rs}_2\ +\
+ \ \epsilon_{rst}\ \eta^{rs}_1 \Phi^{7_s7_t}\eta^{tr}_2 
\eeqa
These zero mode couplings lead to a number of interesting 
4d superpotential couplings for chiral fields. Note that since 
for each E3$^r$-brane instanton there are 8 zero modes, the superpotentials
are going to be quartic in chiral fields, we will not get
directly bilinears. This is due to the presence of the 
$D7$-branes which are required to cancel local twisted
tadpoles.  In particular one gets:
\begin{itemize}

\item
{\bf $\mu$-terms}

Such Higgs mass terms are forbidden perturbatively 
by the $U(1)$ symmetries of the theory. Instantons may generate superpotentials
\beq
W_{\mu}^r\ =\ e^{-S_{E3^r}}\ (1,2,{\bar 2})^r(1,2,{\bar 2})^r\Phi^{7_s7_s}_t\Phi^{7_t7_t}_s\
\ \ ,\ \ r\not=s\not=t\not=r
\eeq
If the singlets $\Phi^{7_s7_s}_t,\Phi^{7_t7_t}_s$ get a non-vanishing 
expectation value, this gives rise to a mass term for the r-th Higgs
multiplet. These vevs may be given without breaking SUSY by switching on 
FI-terms in the $U(1)$'s of the $D7$-branes, which are given by the twisted moduli.
Note that each instanton contributes to a particular Higgs multiplet.
The masses ($\mu$-terms) will depend both on the value of the instanton  action
and on the vevs of the $\Phi^{7_s7_s}_t$ fields.

An alternative to the presence of the 
$\Phi^{77}$ chiral field insertions  is that the compact model was  such that there are 
additional 
projections (e.g. due to Wilson lines on the D7-branes or a $\IZ_2$ Wilson line on the E3-brane instanton) which remove some  zero modes  in the E3-D7 open string 
sector. From the viewpoint of the non-compact example, this would be regarded as the 
non-existence of instanton fermion zero modes in that sector, due to boundary condition at infinity.

\item
{\bf Lepton Yukawa couplings}

In this model perturbative Lepton Yukawa couplings are forbidden
by the $U(1)_L\times U(1)_R$ symmetry (see table \ref{tabpslr}).
Instantons however can generate superpotentials:
\beq
W_Y(leptons)^r\ =\ e^{-S_{E3^r}} 
(1,2,{\bar 2})^r(1,2,1)^s(1,1,{\bar 2})^s \Phi^{7_t7_t}_s \ \ ,\ \ r\not=s\not=t\not=r
\eeq
\beq
W_Y(leptons)'^r\ =\ e^{-S_{E3^r}}
(1,2,{\bar 2})^r(1,2,1)^s(1,1,{\bar 2})^t \Phi^{7_s7_t} \ \ ,\ \ r\not=s\not=t\not=r
\eeq
From eqs.(\ref{trans3},\ref{trans7}) one obtains
that the instanton action transforms like
\beq
S_{E3^r}\ \longrightarrow \ S_{E3^r}\ +\ 2\Lambda_{U(1)_L}\ -\ 2\Lambda_{U(1)_R}
\ +\ \sum_{s\not=r}\ ( \Lambda_{U(1)_1^s}\ -\ \Lambda_{U(1)_2^s})
\eeq
so that the above superpotentials are indeed gauge invariant.

Note that the second superpotential involves non-diagonal $D7^s$-$D7^t$ chiral fields.
As in the case of $\mu $-terms , lepton Yukawa couplings are obtained 
if  fields $\Phi^{7_t7_t}_s$ and/or $\Phi^{7_s7_t}$ get webs, which may be
triggered by $D7$ FI-terms. Note that the flavour structure is then controlled 
both
by the values of the instanton actions as well as the singlet field insertions.
In particular, if one wants to get non-negligible Yukawa couplings the 
exponential suppression should be small and the inserted vevs of singlets large.
On the other hand this has to be done with care because otherwise all three 
Higgs multiplets could  get too large masses from the $\mu$-terms. 
It should be interesting to explore these phenomenological issues further.

\end{itemize}

A couple of comments are in order.
Firstly, in the above discussion
we have  considered the contribution of instantons $E3^r$ with 
$r\not=s$. It is equally easy to compute the
contribution of instantons with $r=s$  but different Chan-Paton 
actions for $E3-$ and $D7-$branes, i.e. for $E3-$ and $D7-$branes wrapping the
same 4-cycle but carrying different CP twist. 
As we mentioned the contribution of those instantons
gives rise to additional fermionic zero modes. In the 
particular model under consideration that would imply some additional insertions 
 of chiral fields  $(D7^rD7^r)_r$
 in the above couplings. 
We have refrained from including those in order to make more
clear the physics.

Secondly, we have assumed above that the E3-instanton only intersects the
D3-branes where the SM sits. Things could however be a bit more tricky upon compactification.
For instance, in simple toroidal orientifolds the E3-brane may be passing in addition through 
the origin, which is fixed under the orientifold and orbifold actions. The local twisted 
tadpoles at that point are $\Tr\gamma_{\theta^k,3}=-4$, $k=1,2$, and require the presence of a 
non-trivial set of D3-branes at such point. The simplest choice is 8 $D3$-branes with 
$\gamma_{\theta,3}=(\alpha \id_4,\alpha^2\id_4)$ and gauge group $U(4)$. Then there would be 
extra fermion zero modes from $D3-E3$ sectors, 4-plets of $U(4)$. This would mean that all
operators discussed above  would be multiplied by factors of the
form $\epsilon_{abcd}6^{ab}6^{cd}$ involving $U(4)$ antisymmetrics.
Thus these operators should also get large vevs if the superpotentials above
are to be present. Again, such vevs can be triggered by the FI-terms associated to the
blow-up modes of these other singularities. We skip this discussion, which would be very model 
dependent, and simply point out that there may exist global compactifications where the E3-brane 
does not intersect other D3-brane sectors, so this potential complication 
 would be absent.

\section{Instantons and non-supersymmetric  $\IC^3/\IZ_N$ 
\\  singularities}
\label{nonsusy}

In the case of  SUSY models the number of universal fermionic zero modes is often larger than 
two, which implies that no non-perturbative charged open string  operators are induced by $E3^r$ 
instantons. As we mentioned in the case of orientifolds the number of zero modes 
is the minimal set of two only for $O(1)$ instantons. This places important constraints on the 
possible non-perturbative effects on these models.

A drastic modification of this problem is to consider 
D3-brane systems at non-SUSY orbifold singularities.
In that case the orbifold projection removes all the universal fermion zero modes from the E3-E3 
sector (in agreement with the fact that, since the geometry does not preserve any supersymmetry, 
there are no Goldstinos on the E3-brane). Thus, there are no extra $\theta^\alpha $ 
fermionic zero modes. It is also important to point out that despite the lack of supersymmetry 
there are E3-branes without tachyonic instabilities, which ensures that they are good saddle 
points of the semiclassical theory.

On the other hand there will be in general charged fermionic
zero modes coming from open strings in the $E3^r-D3$ sectors, coupling to 4d matter scalars.
Integration over those charged zero modes leads to insertions of
4d matter scalar fields, thus generating  non-perturbative scalar interactions. 
Such operators will be generated for any CP structure of the instanton i.e., also for $U(1)$ 
instantons and hence no orientifold projection is required.

Large classes of non-supersymmetric models may be constructed 
locating $D3$-branes at non-SUSY $Z_N$ singularities (see ref.\cite{aiqu}.)
\footnote{Compact non-SUSY orientifolds were discussed in \cite{font}.}.
A subset of those  models are free of tachyons in the closed string sector
\footnote{In particular it is easy to check that
 $Z_N$ singularities with $a_1=a_2=a_3=1$ and $a_4=-3$
(see the appendix for notation) have no tachyons in any twisted sector.
}, and also
in the open string sector, and it is very easy to obtain models
with 3 quark-lepton generations. One expects that $E3^r$ instantons
may yield new non-perturbative couplings in these models.

Let us first summarize some facts about $D3$-branes on non-SUSY
orbifold singularities taken from \cite{aiqu}.
We consider a stuck of $D3$-branes  at a $\IR^6/\Gamma$ singularity with $\Gamma \subset SU(4)$
where, for simplicity,  we take $\Gamma=\IZ_N$.
The $\IZ_N$ action on fermions is given by the  matrix
in eq.(\ref{accionf}) and that on bosons by the matrix (\ref{accionb}).
For the CP twist matrices of D3-branes we consider the general
embedding given by the matrix
\beqa
\label{gamma3}
\gamma_{\theta,3} = \diag (\id_{n_0}, e^{2\pi i/N} \id_{n_1},\ldots,
e^{2\pi i(N-1)/N} \id_{n_{N-1}})
\eeqa
where $\id_{n_i}$ is the $n_i\times n_i$ unit matrix, and $\sum_i n_i=n$.
The  matter spectrum in the $33$ sector is
\beqa
{\rm Vectors} & \prod_{i=0}^{N-1} U(n_i)  \nonumber\\
{\rm Complex} \;\; {\rm Scalars} & \sum_{r=1}^3 \sum_{i=0}^{N-1}
(n_i,{\ov n}_{i-b_r}) \nonumber \\
{\rm Fermions} & \sum_{\alpha=1}^4 \sum_{i=0}^{N-1}  (n_i,{\ov
n}_{i+a_\alpha})
\label{specone}
\eeqa
Note that the spectrum is non-SUSY for $a_4\not=0$. One recovers $N=1$ SUSY 
for $b_1+b_2+b_3=0$, which implies $a_4=0$. 

In general there may be present $D7^r$ branes transverse to the r-th complex plane locally.
There may be CP matrices:
\beq
\gamma_{\theta,7_r}  =  \diag (\ \id_{u_0}, e^{2\pi i/N} \id_{u_1},\ldots,
e^{2\pi i(N-1)/N} \id_{u_{N-1}}) 
\eeq
where we are assuming  $b_r={\rm even}$.  Then one finds matter fields
(for $b_r$ even):
\beqa
\begin{array}{llll}
b_r={\rm even} & \to & {\rm Fermions} &
\sum_{i=0}^{N-1}\, [\, (n_i,{\ov u}_{i+\frac 12 b_r}) +
(u_i,{\ov n}_{i+\frac 12 b_r}) \,] \\
 & & {\rm Complex}\;{\rm Scalars} &
\sum_{i=0}^{N-1}\, [\, (n_i,{\ov u}_{i-\frac 12 (b_s+b_t)}) +
(u_i,{\ov n}_{i-\frac 12 (b_s+b_t)}) \,] \\
\end{array}
\label{spectwo}
\eeqa
with $r\not=s\not=t\not=r$.

We now consider the presence of Euclidean $E3^r$ instantons passing through the 
singularity. As in the SUSY case there will be charged fermionic zero modes
from  $E3^r-D3$ open strings.
Those may be obtained from the corresponding fermionic zero modes
from open strings in the  $D7^r-D3$ sector. Let the CP twist matrix of
the instanton be
\beq
\gamma_{\theta,E3^r} = \diag (\id_{v_0^r}, e^{2\pi i/N} \id_{v_1^r},\ldots,
e^{2\pi i(N-1)/N} \id_{v_{N-1}^r})
\eeq
Then there will be fermionic zero modes from $E3^r-D3$ strings 
transforming like (for $b_r$ even):
\beq
(n_{c-\frac {b_r}{2}}, {\ov v}_c)\ +\ (v_c, {\ov n}_{c+\frac {b_r}{2}})
\eeq
These fermionic zero modes have couplings to scalars 
$\Phi$ in the $D3-D3$ sector
(see eq.(\ref{specone}))
\beq
(n_{c-\frac {b_r}{2}}, {\ov v}_c)\ 
\Phi_{ (n_{c+\frac {b_r}{2}},{\ov n}_{c-\frac {b_r}{2}}) }
\ (v_c, {\ov n}_{c+\frac {b_r}{2}})
\eeq
Integration over the charged fermionic zero modes will give rise to 
determinant couplings among the scalars $\Phi$
of the form
\beq
e^{-S_{E3^r}}\ \times det(  \Phi_{ (n_{c+\frac {b_r}{2}},{\ov n}_{c-\frac {b_r}{2}} )  })
\eeq
Note that this is a purely bosonic coupling. The coupling will be gauge invariant 
due to the $U(1)$ transformation of the euclidean action, as in the SUSY case.
Indeed the derivation of the $U(1)$ gauge transformations described in section 4 
applies also to the case of non-SUSY singularities (and works in full analogy with the 
cancellation of mixed $U(1)$ anomalies in D3/D7-brane systems at non-supersymmetric orbifolds in 
\cite{aiqu}, for reasons already explained).

We thus see that Euclidean $E3$ instantons on this class of non-SUSY Abelian
singularities can give rise to non-perturbative purely bosonic couplings.
Note that the fact that universal fermionic zero modes are projected out
in this class of non-SUSY singularities makes that no fermionic operators
are generated.

In section 3.5 of \cite{aiqu}, an explicit semirealistic model based on a non-SUSY
$Z_5$ singularity is presented. It is a 3-generation left-right symmetric 
model with gauge group $U(3)\times U(2)_L\times U(2)_R\times U(1)^2$
(before some $U(1)$'s get massive by combining with some twisted RR fields).
In that model one can see that e.g. an euclidean E3-brane $U(1)$ instanton with CP matrix = 1
gives rise to a  B-term bilinear in the Higgs multiplet $(2_L,2_R)$, i.e. a term of the form
\beq
e^{-S_{ins}}\ (\Phi_{(2_L,2_R)}\Phi_{(2_L,2_R)}  ) \  \ + \ h.c. .
\label{este}
\eeq
It is interesting to remark that this term gives an example of a
scalar bilinear term which is protected against perturbative loop
corrections without supersymmetry. Indeed, loop corrections
can only give rise to (quadratically divergent) corrections to bilinears of the form
$|\Phi_{(2_L,2_R)}|^2$ but not to terms such as (\ref{este})
which are protected by the perturbatively exact $U(1)$ symmetry. Hence this term
can be hierarchy small compared with the UV cutoff in a completely natural way,
with the small scale generated by the exponential suppression of the instanton.

\section{Final comments}
\label{final}

In this paper we have studied different aspects of the non-perturbative 
superpotentials induced by $E3$ euclidean instantons in systems
with D3/D7-branes sitting at Abelian orbifold singularities.
Much of the recent work on induced superpotentials from
stringy instantons has been formulated in the context of Type IIA
orientifolds with charged matter fields at intersecting 
D6-branes.  The generation of superpotentials in this case 
requires D6-branes wrapping rigid 3-cycles and $O(1)$ orientifold
projections. The construction of globally consistent examples 
within this class has shown to be challenging. 
We have shown how 
in the case
of IIB with D3/D7-branes at singularities finding $E3$ instantons with 
the required fermion zero mode content is much easier.
Furthermore the couplings of the fermionic zero modes to the chiral
$D=4$ fields, which are at  the origin  of 
the  superpotentials,  do not require CFT amplitude
computations but rather come directly given by the singularity
quiver diagrams.
 We have  also sketched
some of the aspects which appear in the case of superpotentials induced
by $E3$ instantons in systems of $D3$-branes sitting
at general toric singularities.

In the systems here studied 
both E3-D3 and E3-D7 fermionic zero modes may contribute
to the amplitudes. The transformation properties of the 
$E3$ instantons under the $U(1)$ symmetries of both $D3$ and $D7$-branes
are nicely compensated by the charges of the 4-D fields
appearing in the induced operator.  
In this way operators perturbatively forbidden by the 
$U(1)$ symmetries are generated by the instantons.

Semirealistic models may be constructed using $D3/D7$ systems
located at Abelian orbifold singularities.
We have shown  how operators with potential phenomenological interest 
can be generated in this context. Some Yukawa couplings 
are often perturbatively forbidden in semirealistic models of
branes at singularities. We have presented examples in which
 forbidden lepton Yukawa couplings are generated due 
to instanton effects. We also presented a global tadpole free  
 $SU(6)$ GUT example  in which u-quark Yukawa couplings are 
generated by these $E3$ euclidean instantons. In a more realistic
three generations  $SU(3)\times SU(2)_L\times SU(2)\times U(1)_{B-L}$
example both a Higgs bilinear ( $\mu$-term) and lepton Yukawa
couplings can be generated. In this example the generation of
these terms will typically require the insertion of vevs for 
$D7-D7$ massless chiral fields.

The examples considered have an unbroken gauged $U(1)_{B-L}$
symmetry which forbids the generation of Majorana neutrino masses,
which is one of the possible interesting applications of 
instanton induced superpotentials. It would be interesting to look
for  semirealistic models in which the $U(1)_{B-L}$ gauge boson becomes
massive by combining with some closed string scalar field so that
Majorana neutrino masses could be generated.

Other possible application of these instanton induced couplings is to
supersymmetry breaking. Indeed, as remarked in \cite{aks} instanton
generated  bilinear couplings combined with $U(1)$ D-terms may lead 
to SUSY breaking a la Fayet  (for fixed closed string moduli).
We have shown an explicit global tadpole free example of this class
based on the $Z_3$ compact orientifold.
More realistic models involving this SUSY breaking mechanism 
should be worth studying. 
On the other hand 
we have found that  certain non-perturbative scalar
couplings are expected to be
generated in the case of systems of $D3$-branes sitting at (tachyon free)
non-SUSY orbifold singularities. This includes the generation
 of exponentially suppressed scalar bilinears which get no
perturbative loop corrections.

One interesting feature of $E3$ instantons that we have described 
is how they can combine with standard gauge instantons to
provide new non-perturbative  superpotentials. We have shown how this
effect described in \cite{geu} can take place even in simple 
toroidal orientifold settings. 
We leave a full exploration of these novel effects for 
future work.

\vspace{50mm}

{\bf Acknowledgments}\\
We thank  G. Aldazabal,
A. Font, I. Garcia-Etxebarria and F. Marchesano, for useful discussions. A.M.U. thanks 
M. Gonz\'alez  for encouragement and support.
This work  has been supported by the European
Commission under RTN European Programs MRTN-CT-2004-503369,
 MRTN-CT-2004-005105, by the CICYT (Spain), the Comunidad de Madrid under project HEPHACOS 
P-ESP-00346 and the Ingenio 2010 CONSOLIDER program CPAN.

\newpage

\centerline{\Large\bf Appendix}

\appendix

\section{Type IIB branes  at Abelian orbifold singularities}
\label{singus}

To fix notation  we  review in this appendix
  the basic formalism to compute the spectrum
and interactions on the world-volume of D3- and
D7-branes at $\IR^6/\IZ_N$
singularities \cite{aiqu}.
Consider a set of
$n$ D3-branes at a $\IR^6/\Gamma$ singularity with $\Gamma \subset SU(4)$
where, for simplicity,  we take $\Gamma=\IZ_N$. Before the projection, the
world-volume field theory on the D3-branes is a $\NN=4$ supersymmetric
$U(n)$ gauge theory.
In terms of component fields, the theory contains $U(n)$ gauge bosons,
four adjoint fermions transforming in the ${\bf 4}$  of the $SU(4)_R$
$\NN=4$ R-symmetry group, and six adjoint real scalar fields transforming
in the ${\bf 6}$.

The $\IZ_N$ action on fermions is given by a matrix
\beq
{\bf R}_{\bf 4} = \diag (e^{2\pi i a_1/N}, e^{2\pi i a_2/N},
e^{2\pi i a_3/N}, e^{2\pi i a_4/N})
\label{accionf}
\eeq
with $a_1+a_2+a_3+a_4=0 \,\, {\rm mod} \,\, N$.
The action of $\IZ_N$ on scalars  can be obtained from the definition of
the action on the ${\bf 4}$, and it is given by the matrix
\beq
{\bf R}_{\bf 6} = \diag (e^{2\pi i b_1/N}, e^{-2\pi i b_1/N},
e^{2\pi i b_2/N}, e^{-2\pi i b_2/N}, e^{2\pi i b_3/N},e^{-2\pi i b_3/N})
\label{accionb}
\eeq
with $b_1=a_2+a_3$, $b_2=a_1+a_3$, $b_3=a_1+a_2$. Scalars can be
complexified, the action on them being then given by ${\bf R}_{esc}=
\diag(e^{2\pi ib_1/N}, e^{2\pi ib_2/N}, e^{2\pi i b_3/N})$.
When $b_1+b_2+b_3=0 $, we have
$a_4=0$ and the $\IZ_N$ action is in $SU(3)$. This case corresponds to a
supersymmetric singularity. The fermions with $\alpha=4$ transforming in
the adjoint representation of $U(n_i)$ become gauginos, while the other
fermions transform in the same bifundamental representations as the
complex scalars. The different fields fill out complete vector and chiral
multiplets of $\NN=1$ supersymmetry.

The   action of the $\IZ_N$ generator $\theta$ will   be embedded on the
Chan-Paton indices. In order to be more specific we consider the general
embedding given by the matrix
\beqa
\label{gamma3}
\gamma_{\theta,3} = \diag (\id_{n_0}, e^{2\pi i/N} \id_{n_1},\ldots,
e^{2\pi i(N-1)/N} \id_{n_{N-1}})
\eeqa
where $\id_{n_i}$ is the $n_i\times n_i$ unit matrix, and $\sum_i n_i=n$.
Analogously for a $D7^r$ which is transverse to the $z^r$ complex
coordinate:
\beqa
\label{gamma7}
\gamma_{\theta,7} = \diag (\id_{u_0}, e^{2\pi i/N} \id_{u_1},\ldots,
e^{2\pi i(N-1)/N} \id_{u_{N-1}})
\eeqa
Then the chiral open string
spectrum in the $N=1$ case with a $D7^r$ transverse to the
$z^r$ complex coordinate is:
\beqa
\begin{array}{cccc}
{\bf 33} & {\rm Vector\,\, mult.} & \prod_{i=0}^{N-1} U(n_i) & \cr
& {\rm Chiral\,\, mult.} & \sum_{i=0}^{N-1} \sum_{s=1}^3 (n_i,{\ov
n}_{i+a_s}) & \cr
{\bf 37_r}, {\bf 7_r 3} & {\rm Chiral\,\, mult.} & \sum_{i=0}^{N-1} \,
[\,(n_i,{\ov u}_{i-\frac 12 a_r}) + (u_i,{\ov n}_{i-\frac 12 a_r}) \, ] &
a_r\; {\rm even} \cr
& & \sum_{i=0}^{N-1} \, [\,(n_i,{\ov u}_{i-\frac 12 (a_r+1)}) + (u_i,{\ov
n}_{i-\frac 12 (a_r+1)}) \, ] & a_r\; {\rm odd}
\label{specsusy}
\end{array}
\eeqa
All chiral fields transform as bifundamentals.
We  denote $\Phi^r_{i,i+a_s}$ the $33$ chiral multiplet in the
representation $(n_i,{\ov n}_{i+a_s})$. We also denote (assuming
$a_r={\rm even}$ for concreteness) $\Phi^{(37_r)}_{i,i-\frac 12 a_r}$,
$\Phi^{(7_r 3)}_{i,i-\frac 12 a_r}$ the $37_r$ and $7_r 3$ chiral
multiplets in the $(n_i,{\ov u}_{i-\frac 12 a_r})$, $(u_i,{\ov n}_{i-\frac
12 a_r})$. With this notation, the interactions are encoded in the
superpotential
\beqa
W & = & \sum_{r,s,t=1}^3\; \epsilon_{rst} \; \Tr (\,\Phi^r_{i,i+a_r}
\Phi^s_{i+a_r,i+a_r+a_s} \Phi^t_{i+a_r+a_s,i} \,) +
\sum_{i=0}^{N-1}\; \Tr (\,\Phi^3_{i,i+a_r}
\Phi^{(37_r)}_{i+a_r,i+\frac 12 a_r} \Phi^{(7_r 3)}_{i+\frac 12 a_r,i} \,)
\label{superp}
\eeqa
There are in general local twisted RR tadpoles. The
conditions for their cancellation is
\beqa
[\, \prod_{r=1}^3 2\sin(\pi kb_r/N)\, ]\; \Tr \gamma_{\theta^k,3} +
\sum_{r=1}^3 2\sin(\pi k b_r/N)\; \Tr \gamma_{\theta^k,7_r} \, = \, 0
\label{tadpogen}
\eeqa

\subsection{The  $\IC^3/\IZ_3$ case }

Most of the examples mentioned  in the main text  make use of this
singularity so that for convenience we summarize here this case.
The   open string chiral  spectrum is given by
\beqa
\begin{array}{cc}
{\bf 33} & U(n_0) \times U(n_1) \times U(n_2) \\
   & 3\, [ (n_0, {\ov n}_1) + (n_1,{\ov n}_2) + (n_2,{\ov n}_0) \, ] \\
{\bf 37_r}, {\bf 7_r3} & (n_0,{\ov u^r}_1) + (n_1, {\ov u^r}_2) +
(n_2,{\ov u^r}_0)+\\
& +(u^r_0,{\ov n}_1) + (u^r_1, {\ov n}_2) + (u^r_2,{\ov n}_0)
\end{array}
\label{spec3}
\eeqa
The superpotential terms are
\beqa
W= \sum_{i=0}^2 \sum_{r,s,t=1}^3 \; \epsilon_{rst} \Tr (\Phi^r_{i,i+1}
\Phi^s_{i+1,i+2} \Phi^t_{i+2,i}) +
\sum_{i=0}^2 \sum_{r=1}^3
\Tr (\Phi^r_{i,i+1} \Phi^{37_r}_{i+1,i+2} \Phi^{7_r3}_{i+2,i})
\label{superpz3}
\eeqa
eeqa
In this  $\IC^3/\IZ_3$ singularity it is possible to consider the generic case
of D7$^{\beta}$-branes, with world-volume defined by $\sum_r \beta_r Y_r=0$,
which preserve the $\NN=1$ supersymmetry of the configuration for
arbitrary complex $\beta_r$. The 37$^\beta$, 7$^\beta$3
spectra are as above, but the superpotential is $W= \sum_i \sum_r \beta_r
\Tr(\Phi^r \Phi^{37^{\beta}} \Phi^{7^{\beta}3})$, with fields from a
single mixed sector coupling to 33 fields from all complex planes.

The twisted tadpole cancellation conditions are
\beqa
\Tr \gamma_{\theta,7_3} - \Tr \gamma_{\theta, 7_1} - \Tr
\gamma_{\theta,7_2} + 3 \Tr \gamma_{\theta,3} = 0
\label{tadpoz3}
\eeqa
These equations are equivalent to the non-abelian
anomaly cancellation conditions.

\section{A $Z_7$ compact example}
\label{zseven}

All compact orientifold examples up to now were  based on the $Z_3$
orientifold. One can see that there are instanton induced
superpotentials in other $Z_N$ examples.
Although it has no phenomenological interest, let us briefly mention the
case of the compact $Z_7$ orientifold which is the next simplest
compact orientifold with only O3-planes.
The twist $\theta $ is generated by $v=\frac {1}{7} (1,2,-3) $.
The twisted tadpole cancellation condition implies
$\Tr \g_{\th}=32\cos\frac{\pi}7 \cos\frac{2\pi}7 \cos\frac{3\pi}7= 4$.
Then locating all D3-branes at the origin with CP matrix
\beq
 \g_{\th} = \diag (\delta \id_4,\delta^2 \id_4, \bar\delta ^3\id_4,
 \id_4,\bar\delta  \id_4, \bar\delta^2  \id_4, \delta^3 \id_4, \id_4)
\eeq
twisted tadpoles cancel.
Here  $\delta ={\rm e}^{2i\pi /7}$ and $\bar \delta = \delta^*$.
The gauge group is $U(4)^3\times SO(8)$
and the  matter spectrum is given by
\beqa
({\underline {4, 1,1}},8_v) &+&({\underline {{\overline 4},
{\overline 4},1 }},1)
+({\underline {6,1,1}},1)      \\
    + ({\ov 4},4,1,1)&+&(1,{\ov 4},4,1)+ (4,1,{\ov 4},1)
\eeqa
where the underlining means one has to add permutations.
Now there are $E3^r$ $O(1)$ instantons transverse to the r-th plane.
They give rise to fermionic zero modes $\eta^r_a$
which transform  like
\beq
\eta^1\ =\ (4,1,1,1)\ ;\ \eta^2\ =\ (1,4,1,1)\ ;\ \eta^3\ =\ (1,1,4,1,1)
\eeq
respectively for $r=1,2,3$ labeling each complex plane.
The main difference here with the case of the $\IZ_3$ orientifold is that
each instanton $E3^r$  has zero modes transforming non-trivially
under a different gauge group.
Each zero mode couples  to a different antisymmetric 6-plet according to
eq.(\ref{inssuperpz3}). These are  couplings of the form
$\eta^r_{a}6^{ab}_r\eta^r_b$.
Then  mass terms  of the form
\beq
\sum_{r=1}^{3}\ e^{-S_{E3^r }}\ \epsilon^{abcd}\ 6_{ab}^r \ 6_{cd}^r\
\eeq
are  generated by the  instanton.
Indeed one can check the $U(1)^3$ gauge invariance of this operator.

\newpage

\section{E3-brane instantons in general toric singularities}
\label{dimers}

In this appendix we provide the basic tools to generalize the analysis in the main text to the computation of non-compact E3-brane instanton effects for systems of D3-branes at general toric singularities. For simplicity (and due to limitations in the available tools) we restrict to configurations with no D7-branes.

Many properties of the geometry of toric singularities, as well as of the gauge theories on D3-branes located at them, can be studied using the so-called brane tilings or dimer diagrams, see e.g. \cite{Hanany:2005ve,Franco:2005rj,Feng:2005gw}. The structure of the gauge theory is encoded in a tiling of a 2-torus by a graph, with faces corresponding to gauge factors, edges to chiral multiplets in bi-fundamental representations (under the gauge factors of the two faces separated by the edge), and nodes to superpotential couplings (among the chiral multiplets associated to the edges ending on the node).
Referring to these papers for details, let us simply say that the corresponding 
gauge theories have a product gauge group $\prod U(n_i)$ and a set of bifundamental chiral multiplets $\Phi_{ij}^{\, a_{ij}}$ in the $(\fund_i,\antifund_j)$, with the index $a_{ij}$ distinguishing possible multiplets in the same representation (in what follows we omit this index $a_{ij}$ for clarity).
We focus on systems of D3-branes at non-orientifold toric singularities, but the 
generalization of our discussion below to systems of D3-branes at orientifolded toric singularities can be treated similarly, using the techniques in \cite{Franco:2007ii}.

We would like to consider possible instanton effects in this kind of configuration. Instantons corresponding to euclidean branes wrapped on the collapsed cycles at the singularity (the analog of E$(-1)$-brane instantons in the orbifolds in the main text), can be efficiently described using the dimer diagrams, and instanton effects for non-gauge instantons (namely euclidean branes associated to a face / gauge factor not occupied by the 4d spacefilling branes) have been described (upon the introduction of orientifold actions) in \cite{Franco:2007ii}. For this reason, and also to keep with the main line in this paper, we rather consider instantons arising from E3-branes wrapped on non-compact 4-cycles passing through the singularity.

The general problem of describing possible non-compact holomorphic 4-cycles in general toric singularities was addressed in appendix B of \cite{Franco:2006es}. The motivation there was to wrap D7-branes on them to introduce flavors for the D3-brane gauge theory, but the results can be applied to the description of the wrapped E3-brane instantons, and of the coupling of their zero modes to the D3-brane fields. 
The main result from the analysis is as follows. 
For each bi-fundamental $\Phi_{ij}$ in the D3-brane gauge theory,
 there exist a non-compact supersymmetric 4-cycle passing through 
the singularity, such that an E3-brane instanton wrapping it leads to 
charged fermion zero modes $\alpha_i$, $\beta_j$, charged in the 
$\fund_i$ and $\antifund_j$ of the $U(n_i)$ and $U(n_j)$ D3-brane gauge factors, and having couplings
\beqa
\Delta S_{E3}\, = \,\alpha_i\, \Phi_{ij} \, \beta_j
\eeqa
with the D3-brane chiral multiplets. When the D3-brane gauge theory has several multiplets in the same gauge representation (labeled by $a_{ij}=1,\ldots, K_{ij}$), they all correspond to E3-branes on the same 4-cycle, but carrying different world-volume gauge bundles (distinguished by a $\IZ_{K_{ij}}$ Wilson line at infinity). This correspondence between 4d bi-fundamentals and non-supersymmetric 4-cycles follows (as discussed in \cite{Franco:2006es}) from the AdS/CFT correspondence between di-baryonic operators and 3-cycles on the 5d horizon of the gravity dual of the D3-brane gauge theory. Namely
the dibaryons are constructed from chiral multiplets by antisymmetrization of 
indices, and the 3-cycles are the bases of 4d cones describing the non-compact 4-cycles.

It is easy to obtain the 4d effective vertex generated by one such D3-brane instanton. 
Let us focus on the case $n_i=n_j$, otherwise the charged fermion zero 
modes are unpaired, leading to a vanishing contribution (or rather, 
requiring the presence of additional intersections of the E3-brane 
instanton with other branes in a globally consistent example, see 
footnote 8). The integration over the charged fermion zero modes leads to a superpotential
\beqa
W\, \simeq \, e^{-T} \, \det \Phi_{ij} 
\eeqa
where $T$ denotes the modulus associated to the 4-cycle in an eventual global embedding of the local configuration. 

It is easy to recover the results for orbifold singularities in this language. Similarly
it is possible to construct explicit models of D3-branes at such more general toric singularities, and to describe the possible effects of E3-brane instantons in the field theory. We refrain from this more extensive discussion, leaving it for future work.

\end{document}